\documentclass[prd,twocolumn,showpacs,nofootinbib,superscriptaddress]{revtex4-1}

\usepackage{amsfonts}
\usepackage{amsmath}
\usepackage{amssymb}
\usepackage{upgreek}
\usepackage{bm}
\usepackage{dcolumn}
\usepackage{epsfig}
\usepackage{graphicx}
\usepackage{graphics}
\usepackage[latin1]{inputenc}
\usepackage{latexsym}
\usepackage{rotating}
\usepackage{float}
\usepackage{hyperref}
\usepackage{xspace} 
\usepackage[usenames,dvipsnames]{color}

\usepackage{ulem}
\definecolor {darkgreen}{rgb}{0.2,0.7,0.2} 

\newcommand\be{\begin{equation}}
\newcommand\ba{\begin{eqnarray}}
\newcommand\ee{\end{equation}}
\newcommand\ea{\end{eqnarray}}
\newcommand{\bes}{\begin{subequations}}
\newcommand{\ees}{\end{subequations}}
\newcommand{\beqn}{\begin{eqnarray*}}
\newcommand{\eeqn}{\end{eqnarray*}}
\newcommand\bw{\begin{widetext}}
\newcommand\ew{\end{widetext}}

\newcommand{\nn}{\nonumber}

\newcommand{\pe}{{\mbox{\tiny PE}}}
\newcommand{\ms}{{\mbox{\tiny MS}}}

\newcommand{\mx}{{\mbox{\tiny max}}}

\begin{document}

\title{Probing the internal composition of neutron stars with gravitational waves}

\author{Katerina Chatziioannou}
\affiliation{Department of Physics, Montana State University, Bozeman, Montana 
59718, USA.}
\author{Kent Yagi}
\affiliation{Department of Physics, Montana State University, Bozeman, Montana 
59718, USA.}
\author{Antoine Klein}
\affiliation{Department of Physics and Astronomy, The University of Mississippi, University, MS 38677, USA.}
\author{Neil Cornish}
\affiliation{Department of Physics, Montana State University, Bozeman, Montana 
59718, USA.}
\author{Nicol\'as Yunes}
\affiliation{Department of Physics, Montana State University, Bozeman, Montana 
59718, USA.}

\date{\today}

\begin{abstract}

Gravitational waves from neutron star binary inspirals contain information about 
the as yet unknown equation of state of supranuclear matter. In the absence of 
definitive experimental evidence that determines the correct equation of state, 
a number of diverse models that give the pressure inside a neutron star as 
function of its density have been constructed by nuclear physicists. These 
models differ not only in the approximations and techniques they employ to solve 
the many-body Schr\"odinger equation, but also in the internal neutron star 
composition they assume. We study whether gravitational wave 
observations of neutron star binaries in quasicircular inspirals up to contact will allow us to distinguish 
between equations of state of differing internal composition, thereby providing important information about the properties and behavior of extremely 
high density matter. We carry out a Bayesian model selection analysis, and find 
that second generation gravitational wave detectors can heavily constrain 
equations of state that contain only quark matter, but hybrid stars containing 
both normal and quark matter are typically harder to distinguish from normal 
matter stars. A gravitational wave detection with a signal-to-noise ratio of 
$20$ and masses around $1.4M_{\odot}$ would provide indications of the existence or absence of strange quark stars, while a signal-to-noise ratio $30$ 
detection could either detect or rule out strange quark stars with a 20 to 1 
confidence.
The presence of kaon condensates or hyperons in neutron star inner 
cores cannot be easily confirmed. For example, for the equations of state studied in this paper,
even a gravitational wave signal 
with a signal-to-noise ratio as high as $60$ would not allow us to claim a 
detection of kaon condensates or hyperons with confidence greater than 5 to 1. 
On the other hand, if kaon condensates and hyperons do not form in neutron 
stars, a gravitational wave signal with similar signal-to-noise ratio would be able to constrain their existence 
with an 11 to 1 confidence for high-mass systems. We, finally, find that combining multiple lower SNR 
detections (stacking) must be handled with caution since it could fail in cases 
where the prior information dominates over new information from the data.

\end{abstract}

\maketitle

\section{Introduction}
\label{intro}

The next generation of gravitational wave (GW) detectors is scheduled to start collecting data in the next months. Initially consisting of Advanced LIGO (aLIGO)~\cite{Harry:2010zz} and Advanced Virgo (AdV)~\cite{Accadia:2012zzb}, 
the network of detectors will be expanded in the next few years to include KAGRA~\cite{Somiya:2011np} and possibly LIGO-India~\cite{ligoindia}. Although the main objective of the advanced detectors is to provide the first direct detection of GWs, the scientific community has started shifting focus to an even more interesting question: Once we have secured the first GW detections, what can we infer about Nature? 

One of the most promising GW sources for extracting physical information is the inspiral of compact binaries consisting of neutron stars (NSs). Not only are they expected to be the most abundant GW source~\cite{2010CQGra..27q3001A}, but also they are well understood. During the $\sim 10^3$ seconds these GWs spend in the detectors' most sensitive frequency range, they can be accurately tracked with theoretical templates. Modeling the detectors' output with these templates cannot only lift the extremely weak signal out of the noise (detection), but also provide us with estimates of the parameters of the GW (parameter estimation). 

One set of parameters that affects the templates is linked to finite-size effects that NSs experience because they are extended bodies with structure. When objects with a finite size are subjected to the tidal field of another object their multipole moments are affected in a way that depends on the equation of state (EoS) -- for barotropic fluids, a relation between pressure and density-- of their matter. The densities encountered in NS interiors are extremely high; in the inner core they even exceed nuclear densities. In this high density regime, laboratory experiments and observations have still to provide a definitive answer on the correct EoS. We, therefore, study whether GWs can be used to answer the following question: Given the detection of the NSNS quasicircular inspiral, can we use finite-size effects to learn about the EoS of the extremely dense NS interior~\cite{Xing:1996sr,Faber:2002zn,Faber:2003sb,GondekRosiska2007271,lattimer-prakash-review,Flanagan:2007ix,Hinderer:2009ca,PhysRevD.85.123007}? 

To leading order, finite-size effects cause the quadrupole moment tensor of a star 
$Q_{ij}$ to be affected by the tidal field tensor of its companion ${\cal{E}}_{ij}$ 
through $Q_{ij}=- \lambda {\cal{E}}_{ij}$. The constant of proportionality 
$\lambda$ is called the tidal deformability and it is a function of the mass and the 
EoS. This tidal interaction causes NSs to be distorted during the inspiral phase 
and torn apart before merger~\cite{1992ApJ...400..175B}. By the `inspiral phase' we define the evolution of the binary up to an orbital separation of six times the total mass or contact, whichever comes first. The plunge and final collision after this orbital separation is called the `merger phase'.
Since this merger is 
expected to happen at high frequencies of ${\cal{O}}(10^3)$Hz, where the 
detector noise is likely to dominate, we here focus on the better modeled 
inspiral part{\footnote{A number of studies have examined the possibility of 
determining the EoS from the merger phase (see~\cite{Lackey:2013axa} for an 
example). However, in the absence of a full and accurate template bank of merger 
waveforms, it is not clear how one would perform a full data analysis study.}}. 
The relatively small velocities (never exceeding ~0.3 times the speed of light) 
of this inspiral phase make it ideal for a \emph{post-Newtonian} (PN)
description\footnote{A term of relative order $(v/c)^{2N}$ is said to be a N-PN 
order term.} in most of the frequency range considered, where all quantities are expanded in powers of 
$v/c$~\cite{Blanchet:2014zz}.  The PN waveform, in fact, becomes less accurate near the merger~\cite{Favata:2013rwa,Yagi:2013baa,Wade:2014vqa}. For nonspinning NS binaries, a more accurate waveform is available using the effective-one-body (EOB) approach~\cite{Bernuzzi:2014owa}. In this paper, we consider precessing NS binaries as we will explain in more detail below. Since the EOB waveform for precessing NS binaries is currently unavailable, we use the precessing PN waveform.

In the PN framework, the first finite-size effect enters the waveform at 2PN order in the GW phase through spin corrections to the quadrupole moment of the objects $Q_{1,2}$~\cite{PhysRevD.57.5287}, and here we also include the 1PN correction to this~\cite{spin-ilq}. Then, at 5PN order and above, the tidal deformabilities $\lambda_{1,2}$ enter the phase directly~\cite{Flanagan:2007ix,Vines:2011ud,Hinderer:2009ca,PhysRevD.85.123007,Favata:2013rwa,Yagi:2013baa}. All these parameters are EoS-dependent, however, two of the authors showed that their interrelation is approximately EoS-independent~\cite{Yagi:2013bca,Yagi:2013awa}. A lot of work has been put into understanding and extending this result~\cite{lattimer-lim,maselli,I-Love-Q-B,Doneva:2013rha,Pappas:2013naa,Chakrabarti:2013tca,Baubock:2013gna,Sham:2013cya,Yagi:2014qua,Doneva:2014faa,Pani:2014jra,Kleihaus:2014lba,Martinon:2014uua,Chan:2014tva,Yagi:2015hda,Yagi:2015upa,Majumder:2015kfa,Delsate:2015wia}, in particular to include higher-order multipole moments and tidal deformabilities~\cite{Pappas:2013naa,Stein:2013ofa,Yagi:2013sva,Chatziioannou:2014tha,Yagi:2014bxa}. We can therefore use the Love-Q relation to choose $Q_{1,2}$ in favor of $\lambda_{1,2}$ or vice versa from the GW phase.

The problem of the detectability of finite size effects with gravitational waves has gathered a lot of attention in recent years. Initial studies, based on quantifying the differences between waveforms with different EoSs or on a Fisher information matrix analysis~\cite{Read:2009yp,Hinderer:2009ca,Pannarale:2011pk,PhysRevD.85.123007,PhysRevD.85.044061,Read:2013zra,Maselli:2013rza,Lackey:2013axa} suggested that aLIGO has the potential of providing useful information on the NS EoS. However, due to the expected low signal-to-noise ratio (SNR) in aLIGO detections and due to the strong correlations between the different GW parameters, the applicability of a Fisher study, and the conclusions derived from it, is limited~\cite{Vallisneri:2011ts,Rodriguez:2013mla,Cho:2012ed,O'Shaughnessy:2013vma}.

For this reason, several Markov-Chain Monte-Carlo (MCMC)~\cite{Cornish:2007ifz,Littenberg:2009bm} studies have recently been carried out in order to address EoS detectability in a more robust way in the context of Bayesian inference. The first such study was performed by Del Pozzo et al.~\cite{DelPozzo:2013ala}, which showed that a few tens of detections of moderate brightness can be combined to provide strong constraints on the EoS, though this result seems to be highly dependent on the mass distribution of the sources~\cite{Agathos:2015uaa}. Wade et al.~\cite{Wade:2014vqa} studied the effects of systematic and statistical errors in EoS extraction, while Lackey and Wade~\cite{Lackey:2014fwa} employed a more realistic parametrization of the EoS and agreed that a few bright sources can determine the EoS of NSs. 

With the exception of~\cite{DelPozzo:2013ala} and~\cite{Agathos:2015uaa}, all previous work mentioned above consisted of \emph{parameter estimation studies}, where the tidal deformability is treated as a system parameter and searched over with an MCMC analysis. Any EoS that predicts a value of the deformability within the recovered uncertainty is compatible with the results of the MCMC. Any EoS that does not fall in the deformability error bars can be ruled out.  We choose a more direct approach here and compare the different EoS models directly, an approach known as \emph{model selection}. In the latter, the EoS itself, instead of the tidal deformability is treated as an independent parameter of the system and the analysis allows the data to select which EoS is preferred. Our study is unique in that we use this tool to perform a comprehensive study on whether \emph{we can extract important physical information about the composition of NSs, such as the existence of exotic species}. 

The different EoS models proposed in the literature differ not only in the NS composition they assume, but also in the approximate schemes they employ to solve the many body Schr\"odinger equation. These approximate schemes include approaches such as the variational method~\cite{Pandharipande:1971up}, Skyrme-Hartree-Fock (SHF) models~\cite{1956PMag....1.1043S}, Brueckner-Hartree-Fock (BHF) models~\cite{1996ApJ...469..794E}, and relativistic mean field (RMF) theory~\cite{Shen1,Shen2}. The NS internal composition might be that the EoS is constructed solely with normal matter (neutrons n, protons p, electrons e, muons $\mu$), or it may contain kaon condensates (K), hyperons (H), pion condensates ($\pi$), or quark matter (Q).

Determining that $2$ EoSs with the \emph{same} composition but \emph{different} approximations are distinguishable will not result in any new information about Nature. On the other hand, determining that $2$ EoSs with \emph{different} compositions and \emph{different} approximations are distinguishable must be treated with caution. Can we claim that the difference between the EoSs we detected is due to their actual physical differences, or due to their distinct mathematical approaches? In order to avoid this obstacle, we compare EoSs that employ the \emph{same}, or as similar as possible, approximations and differ only in their internal composition.  

For example, consider an EoS, which contains normal matter, and is constructed with the variational method{\footnote{Refer to Appendix~\ref{app-eos} for a description of all EoSs used here along with their physical content and approximation schemes.}}, and another EoS, which is also constructed with the variational method, but includes both normal matter and hyperons. When we perform a comparison on these $2$ EoSs, we are essentially comparing a hyperon EoS and a hyperonless EoS. The result of this comparison can directly be translated to physical information. If the hyperon EoS is preferred, we have detected hyperons in NS cores assuming one of the 2 EoSs is correct. If the hyperonless EoS is preferred, we have constrained hyperons in NS cores.

At this stage one may reasonably object: Could we confuse a hyperon EoS constructed with one method with a normal matter EoS constructed with a different method? This concern can be alleviated by performing a large number of comparisons. Concluding that a hyperon EoS constructed with one method is preferred over a normal matter EoS with the same method is not enough. We need to find as many pairs of EoSs that have been constructed with a variety of different methods and compare all of them. This approach (i) ensures we have not confused the effects of internal composition and approximation schemes, and (ii) provides us with insight on how hyperons, or other particles, affect EoSs in general. Effectively, by comparing many pairs of EoSs, we `average out' any effect coming from how each pair of EoSs is constructed and isolate the effect of the common difference between the 2 EoSs of all pairs i.e.~the exotic matter.

Another aspect in which our analysis generalizes previous studies is that for the first time we include spin-precession both in our simulated signal and in the templates. We use the fully analytic double-precession model of~\cite{Chatziioannou:2013dza}, derived under the assumption that the binary components have small spins. This turns out to be an excellent approximation for NSs in the LIGO band, as their dimensionless spin parameter (the spin angular momentum over the mass squared)  is not expected to exceed $\sim 0.1$~\cite{Mandel:2009nx}. The model has already been tested before in a data analysis context~\cite{Chatziioannou:2014coa,Chatziioannou:2014bma}, however, in those studies we stopped our analysis at $400$Hz to avoid finite-size effects~\cite{Lai:1996sv}, and focus instead on the measurability of the masses and the spins.

The inclusion of spin-precession in the templates is crucial. Reference~\cite{Chatziioannou:2014coa} showed that allowing for the systems to precess around the orbital angular momentum can break degeneracies between the masses and the spins, improving mass extraction by about an order of magnitude. This improved mass determination is directly translated to better $\lambda$ extraction, making EoSs easier to distinguish. This is because in the context of model selection, it is the EoS that is a GW parameter, and not $\lambda$, which is determined through a relation of the form $\lambda(m,\text{EoS})$ (see Sec.~\ref{models}). We should emphasize an important distinction here: the effect of better mass extraction due to precession has nothing to do with the actual spin magnitude of the signal we study. What is important, however, is allowing for the template we recover the signal with to model precessional effects.

The main results of our analysis are summarized below.

\emph{We find that advanced detectors will be able to place strong constraints on the existence of quark stars comprised solely of quark matter.} Given the EoSs available today, a NSNS binary with masses in the $(1.2,1.5)M_{\odot}$ range with SNR $=30$ can effectively rule out strange quark stars, or make a positive detection of them. We, furthermore, argue that even for the plausible EoSs constructed in the future, there exists some mass $\in(1M_{\odot},1.8M_{\odot})$ where quark stars are distinguishable from normal matter NSs.

\emph{The prospects of detecting or ruling out hybrid NSs including both normal and quark matter are worse.} If the strong interactions between quarks are close to those predicted by a perturbative analysis~\cite{Fraga:2001id} and the transition between nuclear and quark matter phases happens around twice the nuclear saturation density, the detection of a $(1.4,1.35)M_{\odot}$ NS binary with SNR $30-40$ could provide significant evidence of whether quarks form in NS interiors. However, if the strong interactions are weaker, aLIGO will not be able to reach confident conclusions.

\emph{It is unlikely that aLIGO will be able to claim a detection of hyperons or 
kaons, since that would require high masses and SNR $\gtrsim 60$.} The detection 
of hyperon or kaon condensates in NS interiors requires high mass stars, since 
it is only at these high masses that you encounter densities large enough for 
these condensates to form. This poses a significant problem; the importance of 
finite size effects is reduced with increasing mass since $\lambda$ decreases 
with increasing mass. Moreover, most NSs are expected to have masses around 
$\sim 1.4M_{\odot}$, rarely reaching the $2M_{\odot}$ required for hyperons and 
kaons detection. We therefore conclude that it is unlikely that aLIGO will be able to 
positively identify hyperons or kaons in NSs. On the other hand, if hyperons and 
kaons are \emph{not} formed in NS interiors, aLIGO could place constraints on 
their existence. 

\emph{Our analysis suggests that aLIGO can distinguish between models that differ at low central densities, like normal matter EoSs and EoSs containing quark matter.} In order to probe the high density regime we need SNRs higher than what aLIGO is likely to achieve. This is due to the fact that high mass systems (i) present smaller finite size effects, and (ii) have masses close to the maximum mass allowed, which causes some interesting effects related to the prior boundary (see Sec.~\ref{stacking}).

\emph{Among the various noise configurations aLIGO can be tuned to, the optimal for EoS determination is the default Zero-Detuned, High-Power (Zero-Det., High-P) one~\cite{AdvLIGO-noise}.} Tuned configurations include the NSNS Optimized (NSNS Opt.) and the High Frequency (High F.) ones; both perform in an inferior way when it comes to EoS extraction. In the case of NSNS Opt.~this is due to its low sensitivity at frequencies above $600$Hz, when it is exactly at these frequencies that finite size effects are more prominent. As far as High F. is concerned, its improved sensitivity is limited to a very narrow frequency range around $10^3$Hz. This fact coupled to its worsened sensitivity at low frequencies makes High F. unsuitable for EoS studies. We conclude that in the high frequency regime we are interested in, the Zero-Det., High-P.~noise curve has the overall higher sensitivity and more accumulated SNR. 

\emph{Stacking, i.e.~combining multiple low SNR sources, might improve the results obtained here, but it might also lead us to erroneous conclusions about the true EoS.} We find that when the SNR is low our results can be dominated by prior information, rather than any new information we get from the GW data. This could lead to each individual low SNR binary system providing some confidence, but in favor of the wrong EoS. Stacking all these systems will inevitably lead to great confidence in favor of the wrong EoS. We emphasize that stacking must always be treated with caution.

\emph{As far as the spin of the bodies is concerned, we find that the magnitude of the injected spin angular momentum has a negligible effect on our results, provided that the templates allow for spin precession.} This is in agreement with Ref.~\cite{Chatziioannou:2014coa}, where the order-of-magnitude improvement in mass extraction was achieved over spin-aligned templates even for nonspinning systems, as long as spin-precessing templates were used in the recovery of the signal.

The remainder of the paper is organized as follows. 
In Sec.~\ref{ms}, we describe the techniques and simulations we use. 
In Sec.~\ref{EOS}, we describe in more detail the results of the EoS comparison.
In Sec.~\ref{discussion}, we summarize our work.
Throughout the paper we use units where $G=c=1$.

\section{Model Selection}
\label{ms}

Model selection in the Bayesian framework requires an explicit statement of the models compared. In this section, we describe in detail the models we use, as well as our methodology when comparing them. We describe the \emph{Bayes factor} (BF), a quantity that assesses which model is preferred by the data, and give an overview of our simulated signals. We conclude this section with a brief discussion of the power of stacking signals versus detecting a single loud signal.

\subsection{Bayesian inference}

In the context of Bayesian inference, the probability that a hypothesis $H_1$ is correct given some data $d$ is~\cite{Cornish:2007ifz,Littenberg:2009bm}
\be
p(H_1|d) = \frac{p(H_1) p(d|H_1)}{p(d)},
\ee
where $p(H_1|d)$ is the posterior belief in the hypothesis after the data has been analyzed, $p(H_1)$ is the prior belief based on all information we have before analyzing the data, and $p(d)$ is the probability of the data, an unimportant normalization constant in our case. The evidence $p(d|H_1)$ is given by an integral over the parameters of the model $\bm{\theta}$
\be
p(d|H_1)=\int d \bm{\theta} \; p(\bm{\theta}|H_1) \; p(d| \bm{\theta} H_1),\label{evidence}
\ee
where $p(\bm{\theta}|H_1)$ is the prior information on the model parameters, and $p(d| \bm{\theta} H_1)$ is the likelihood, where $\ln p(d| \bm{\theta} H_1) = -1/2 \left(\left.s-h\right|s-h\right)$ in Gaussian noise, with $s$ the signal, $h$ the template model and $\left(\left.\cdot\right|\cdot\right)$ the noise-weighted inner-product~\cite{Cornish:2007ifz,Littenberg:2009bm}. 

When we have to select between 2 competing hypotheses, we compare their posterior beliefs through the \emph{odds ratio} (OR), defined by
\be
{\cal{O}} = \frac{p(H_1|d)}{p(H_2|d)}=\frac{p(H_1) p(d|H_1)}{p(H_2) p(d|H_2)}.
\ee
The OR is the `betting odds' of $H_1$ compared to $H_2$ and includes the both prior belief in each hypothesis and any new information that is extracted from the data. These two contributions can be separated by defining the \emph{Bayes factor} (BF) 
\be
\text{BF}= \frac{p(d|H_1)}{p(d|H_2)},\label{bf}
\ee
which includes only the data contribution to the OR. A BF $>1$ means that $H_1$ is supported better by the data, while a BF $<1$ means that $H_2$ is preferred. In the case of uninformative priors, i.e.~$p(H_1)=p(H_2)$, the OR equals the BF. In this paper, we choose to work with the BF instead of the OR because we are interested in whether the data lends more support to some hypothesis over another, irrespective of our prior beliefs in them. Of course, once we can confidently quantify our prior belief in a hypothesis, we can trivially go from the BF to the OR.

Working with the BF, however, has one major problem: we cannot draw the same conclusions about two different pairs of hypotheses that have the same BF. For example, consider the problem of whether a given GW signal is better described by GR or by a modified gravity theory. In this case, we have a strong prior belief in favor of GR, given the many successes of experimental relativity~\cite{Will:2014xja,Yunes:2013dva}. We would therefore require very large BFs in favor of the modified gravity hypothesis to claim a detection of a deviation from GR. On the other hand, if we are interested in whether the signal is better described by one of two competing EoSs, we do not have such strong prior beliefs in favor of any of the models. That means that we would not need such large BFs to claim that we have identified the correct EoS of Nature, provided we remain agnostic about the two EoSs.

For this reason, it is crucial that we explicitly define what we mean by ``a BF that is large enough'' on a problem by problem basis. As we will discuss later in this section, we work with models for which we do not have strong experimental prior knowledge. Therefore, we adopt the \emph{Jeffreys scale of interpretation of BFs}~\cite{BF} to define how significant a BF is. When BF $< 1$ it is \emph{negative}, for $1<$ BF $<3$ it is \emph{barely worth mentioning}, for $3<$ BF $< 10$ it is \emph{strong}, for $10<$ BF $<100$ it is \emph{very strong}, and finally for BF $>100$ it is \emph{decisive}. This is in contrast to the analysis of Refs.~\cite{Cornish:2011ys,Sampson:2013lpa,Sampson:2013jpa,Sampson:2014qqa}, which dealt with tests of GR, where BFs around 100 were considered strong and not decisive, given the strong prior in favor of GR.

In order to calculate the BF between two models, a number of different techniques can be used: thermodynamic integration~\cite{Littenberg:2009bm,Veitch:2014wba}, nested sampling~\cite{Veitch:2014wba} and reverse jump, Markov chain-Monte Carlo~\cite{Cornish:2007ifz}. Here, we employ the third technique which requires promoting the model to a parameter of the Markov chains. Then, the BF is given by
\be
\rm{BF}=\frac{\text{time the chains spend in model 1}}{\text{time the chains spend in model 2}}\label{bf-def},
\ee
with error bars calculated with the technique suggested in~\cite{Cornish:2014kda}.

\subsection{Models}
\label{models}

Our models represent GWs emitted in the late inspiral of NS binaries assuming GR is correct from the time they enter the LIGO band until the NSs come into contact, or their separation becomes six times the total mass, whichever comes first (see~\ref{inj} for details). The difference in the models will only be in the finite size effects they include. Comparison between these models can be viewed as traditional parameter estimation, only now the model itself is an extra discrete parameter. In parameter estimation, the waveform of a NS binary inspiral with a circular orbit depends on the following parameters\footnote{In this argument, we encode the EoS dependence through the dimensionless tidal deformabilities $\bar{\lambda}_{1,2}$; the argument goes through if one uses any other quantity to parametrize tidal deformations.}: 
\begin{align}
\bm{\theta}_{\pe}=\{&m_1,m_2,\theta_N,\phi_N,D_L,\theta_L,\phi_L,t_c,\phi_c,\bm{S}_1,\bm{S}_2,\bar{\lambda}_1, \bar{\lambda}_2\},
\end{align}
while in a model selection study the template depends on the parameters 
\be
\bm{\theta}_{\ms}=\{m_1,m_2,\theta_N,\phi_N,D_L,\theta_L,\phi_L,t_c,\phi_c,\bm{S}_1,\bm{S}_2,\rm{EoS} \},
\ee
where $m_i$ are the component masses, $(\theta_N, \phi_N)$ give the sky location of the source, $D_L$ is the luminosity distance, $(\theta_L,\phi_L)$ give the direction of the orbital angular momentum, $(t_c, \phi_c)$ are the time and phase of coalescence respectively, $\bm{S}_i \equiv \chi_i m_i^2 (\sin{\theta_i} \cos{\phi_i},\sin{\theta_i} \sin{\phi_i},\cos{\theta_i})$ are the spin angular momentum vectors of each binary component, with $\chi_{i} := |{\bm{S}}_{i}|/m_{i}^{2}$ the dimensionless spin parameter, and $\bar{\lambda}_i \equiv \lambda_i/m_i^5$ are the dimensionless tidal deformabilities. The two parameter sets are equivalent, since knowledge of $(m_1, m_2, \rm{EoS})$ from $\bm{\theta}_{\ms}$ can be used to construct the quantities $\bar{\lambda}_1(m_1,\rm{EoS})$ and $\bar{\lambda}_2(m_2,\rm{EoS})$ in $\bm{\theta}_{\pe}$. 

The EoS is what \emph{defines} a model. For some EoS the model predicts a GW with $\bm{\theta}_{\pe}$ and $\bar{\lambda}_{1,2}$ as a function of $(m_{1,2},\text{EoS})$ for $(m_1, m_2) \leq M_{\mx}(\text{EoS})$, and no GW otherwise. The quantity $M_{\mx}(\text{EoS})$ is the maximum NS mass that can be stably supported, given an EoS. For the GW template itself, we will use the small-spins, double-precessing waveform of~\cite{Chatziioannou:2013dza}, which is constructed in the Fourier domain through a stationary-phase approximation in a post-Newtonian expansion. We use an extended version of these templates by adding finite-size effects at 2 and 3PN order (due to the quadrupole moment)~\cite{PhysRevD.57.5287,spin-ilq} and at 5 PN order and higher (due to tidal deformability effects)~\cite{Flanagan:2007ix,Vines:2011ud,Hinderer:2009ca,PhysRevD.85.123007}. More specifically, we include tidal terms that depend on the $\ell=2$ electric tidal deformability $\lambda$ up to 7.5PN order given in~\cite{PhysRevD.85.123007}. We also include the contribution from the $\ell =2$ magnetic tidal deformability $\sigma_2$ at 6PN order and the $\ell=3$ electric tidal deformability $\lambda_3$ at 7PN order~\cite{Yagi:2013sva}. We further take into account the leading correction to the adiabatic approximation entering at 8PN order that depends on the $\ell=2$ f-mode frequency $f_2$ of a NS~\cite{Flanagan:2007ix,Hinderer:2009ca}. $\sigma_2$, $\lambda_3$ and $f_2$ can approximately be expressed in terms of $\lambda$ thanks to the universal relations found in~\cite{Yagi:2013sva,Chan:2014kua}. In addition, we use the Love-Q relations~\cite{Yagi:2013bca,Yagi:2013awa} to rewrite the quadrupole moment in terms of the tidal deformability. 

\begin{table}
\begin{centering}
\begin{tabular}{c|c}
\hline
\hline
\noalign{\smallskip}
EoS & Composition \\
\hline
\noalign{\smallskip}
AP4~\cite{APR}, SV~\cite{Lim:2013tqa}, SGI~\cite{Lim:2013tqa}, SkI4~\cite{1995NuPhA.584..467R} & \\
DBHF$^{(2)}$(A)~\cite{2014arXiv1410.7166K}, MPa~\cite{2014PhRvC..90d5805Y}, G4~\cite{Lackey:2005tk} & n, p, e, $\mu$ \\
GA-FSU2.1~\cite{2013JPhG...40b5203G}, Shen~\cite{Shen1,Shen2} & \\
\noalign{\smallskip}
\hline
\noalign{\smallskip}
SGI-YBZ6-S$\Lambda$$\Lambda$3~\cite{2014arXiv1412.5722L}, NlY5KK$^{*}$~\cite{2014arXiv1410.7166K}, & \\
SkI4-YBZ6-S$\Lambda$$\Lambda$3~\cite{2014arXiv1412.5722L} & n, p, e, $\mu$, H \\
MPaH~\cite{2014PhRvC..90d5805Y}, H4~\cite{Lackey:2005tk}  & \\
\noalign{\smallskip}
\hline
\noalign{\smallskip}
SGI178~\cite{Lim:2013tqa} & \\
SV222~\cite{Lim:2013tqa} & n, p, e, $\mu$, K \\
GA-FSU2.1-180~\cite{2013JPhG...40b5203G} & \\
\noalign{\smallskip}
\hline
\noalign{\smallskip}
ALF4~\cite{Alford:2004pf}, ALF5 & n, p, e, $\mu$, $\pi$, Q \\
\noalign{\smallskip}
\hline
\noalign{\smallskip}
SQM3~\cite{SQM} & Q (u, d, s) \\
\noalign{\smallskip}
\hline
\hline
\end{tabular}
\end{centering}
\caption{Classification of EoSs with respect to internal composition. The first cluster corresponds to EoSs with normal matter. The second and third clusters include hyperons and kaon condensates respectively. The last two rows list EoSs that include quark matter.}
\label{table:EoS-super-summary}
\end{table}
Which EoS models should we allow the data to select from? There is a great number of EoSs available in the literature, varying both in the type of matter they consider (quarks, hyperons, kaons, muons, pions, neutrons, protons, electrons), and in the approximation schemes used to construct the EoS (see Appendix~\ref{app-eos}). We can classify the EoSs by the type of matter they employ; the subset of EoSs discussed in this paper are presented in Table~\ref{table:EoS-super-summary}. The EoSs in each of these categories differ in the approximation schemes used to solve the many-body Schr\"odinger equation (see Appendix~\ref{app-eos}).

\begin{figure}[t]
\begin{center}
\includegraphics[width=\columnwidth,clip]{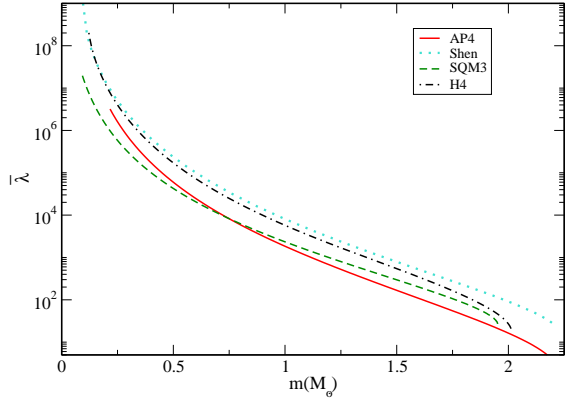}
\caption{\label{fig:love-m_all}~Dimensionless tidal deformability as a function of mass for a number of EoSs with very different physical contents. The lack of clustering in the  $\bar{\lambda}-m$ space shows that we cannot perform model selection with all EoSs simultaneously. }
\end{center}
\end{figure}

Given the great variety of EoSs, how should we carry out a model selection study? Could we select a characteristic set of models, perhaps chosen by looking at how the $\bar{\lambda}-m$ relations behave for a set of EoSs? Figure~\ref{fig:love-m_all} shows these relations for a few of the EoS listed above. Notice how the normal matter EoSs (AP4, Shen) form a band in the $\bar{\lambda}-m$ space that contains both SQM3 (consists of pure quark matter) and H4 (contains hyperons). This \emph{lack of clustering} with respect to the internal composition classification in the $\bar{\lambda}-m$ space indicates that comparing arbitrary EoSs will not produce physically meaningful results. For example, say we concluded that AP4 (red solid line) is distinguishable from H4 (black, dot-dashed line). Can we claim that hyperons are detectable from GW observations? The answer is no; there is another normal matter EoS (Shen, turquoise dotted line) much closer to H4 than AP4. Unless we compare Shen to H4 too, we cannot claim detectability of hyperons. 

Another possibility would be to compare each model against every other model \emph{simultaneously}, but (i) the computational cost would be prohibitive, and (ii) it is not clear what physical question such a comparison would address. Comparing models within each category that differ only in the approximation scheme used to calculate the EoS is analogous to comparing post-Newtonian models for the GWs emitted in the inspiral of compact objects. In the end, all of such models are only approximations to the exact solution of Nature. Instead, we are interested in specific physical questions regarding the internal composition of NSs. To do so, we will choose pairs of EoSs that are as similar as possible (using the same approximation schemes), but differ only in the matter degrees of freedom. In this way, we can directly translate \emph{distinguishability between models} into \emph{distinguishability between physical scenarios}.

\subsection{Signal injections}
\label{inj}

Whether the data can distinguish a given EoS model over another depends strongly on the particular signal detected. The parameters that affect EoS distinguishability the most are the two masses, the distance to the source, which effectively controls the SNR, and the EoS itself. We therefore simulate different signals with various values for these parameters.  For the remaining parameters we select the following injection values: $(\cos{\theta_N}, \phi_N) = (-0.105,3.705)$, $(\cos{\theta_L}, \phi_L) = (0.801,3.216)$, $(t_c, \phi_c) = (1,024 \, {\rm{s}},4.461)$, $(\cos{\theta_1},\phi_1) = (0.774,2.248)$, $(\cos{\theta_2}, \phi_2) = (0.968,5.311)$, and $(\chi_1, \chi_2) = (0.04,0.04)$. All parameters have been randomly chosen so that they do not lead to any `special' orientation of the binary (face on, edge on etc.). We have also performed simulations with other dimensionless spin magnitudes and found that the spin has a very small effect on EoS distinguishability. 

The initial misalignment of the spin and the orbital angular momenta means that the system will undergo precession~\cite{Apostolatos:1994mx}. Indeed, with these choices of parameters, the angle between the orbital angular momentum and the total spin angular momentum is $\sim 30^{\circ}$. Spin precession is modeled through the small-spins, double-precession approach of~\cite{Chatziioannou:2013dza}, which has been shown to be highly accurate for modeling NS binaries~\cite{Chatziioannou:2014bma}. This approach is valid in the inspiral phase only, since it is based on a post-Newtonian expansion. The signal is thus modeled with the same approach as the templates that define the EoS models. 

The approximations used to describe the orbital motion in the small-spins, 
double-precession approach are only valid up to a given frequency. We thus carry 
out our analysis up to $\min(f_{\text{ISCO}},f_{\text{c}}(\text{EoS}))$, where 
$f_{\text{ISCO}}$ is the Keplerian frequency at $r=6 
(m_{1}+m_{2})$~\cite{Uryu:2000dw,Bejger:2004zx} and $f_{\text{c}}$ is the 
\emph{contact frequency}, that at which the separation of the two bodies is 
equal to the sum of their radii. Although Mandel et al.~\cite{Mandel:2014tca} 
showed that terminating the waveforms at a certain frequency can affect the 
results through the addition of artificial information, our cutoffs are at such 
high frequencies that they are not expected to affect the results.

Even though the particular noise realization in the detector at the time the GW 
passes through will have an effect on parameter extraction, we do not inject noise in our analysis. Given that it is 
impossible to predict the noise instance, the best we can do is average our results over multiple noise 
realizations. However, Nissanke et al.~\cite{2010ApJ...725..496N} showed that 
such averaging is equivalent to zero injected noise in 
the signal. Sampson et al.~\cite{Sampson:2013lpa} showed that a given 
noise realization causes the likelihood to shift as a whole, without 
significantly changing its shape (Fig.~4 of~\cite{Sampson:2013lpa}), suggesting that the integral of the likelihood (the evidence) is minimally affected by noise fluctuations. This picture, however, does not hold for poorly constrained parameters, like the ones studied here, where the posterior extends over a large fraction of the prior volume~\cite{Wade:2014vqa,Vallisneri:2011ts}.

Of course, the noise curve of the detector does play a very important role in the calculation of BFs, through the noise-weighted inner product in the likelihood~\cite{Cornish:2007ifz}. We will mostly adopt the zero-detuned, high-power noise spectral density of the advanced detectors~\cite{AdvLIGO-noise}, though we will explore other choices in Sec.~\ref{noise}. The specific form of the likelihood we use assumes that the noise in the detectors is stationary and Gaussian [see below Eq.~\eqref{evidence}], neither of which is strictly true. Cornish and Littenberg, however, have shown how to model the non-Gaussian features~\cite{Cornish:2014kda}, and deal with the nonstationarity of the noise~\cite{Littenberg:2014oda}, leaving us with only the modeling of the stationary, Gaussian noise component. 

When recovering the parameters, we use a uniform prior in $(0.1,3)M_{\odot}$ for the masses, a uniform prior on the sphere for all directions, a uniform prior in $(0,1)$ for the dimensionless spin magnitudes, and a uniform prior in the log of the distance. All prior ranges are selected such that they are wide enough to not affect our results.

\subsection{Stacking vs high SNR}
\label{stacking}

In reality aLIGO will probably reach physically interesting conclusions by combining information from multiple detections, rather than by waiting for a very loud one. Along those lines, one would argue it makes more sense to stack a sufficient number of moderate SNR sources rather than study the BF as a function of the SNR. Our results, however, suggest that stacking should be performed with caution.

For low SNR detections we find $2$ rather counterintuitive effects: (i) it is possible for the wrong model to be preferred over the correct one, and (ii) it is possible for the correct model to be preferred less and less as the SNR increases (see Appendix~\ref{cutoffs}). The first effect is not new; it has already been encountered in the context of comparing models with different dimensionality~\cite{Cornish:2011ys}, where it takes the form of an \emph{Occam Penalty} on the more complicated model. The second effect is perhaps less familiar: why would it be that, as the signal strength increases, the data fails to increasingly support the correct model? The answer to this question and the root cause of these effects can be traced down to sharp cutoffs of the prior distribution.

When computing BFs, we must compare models with different maximum masses. In fact, most of the time the difference is rather large, with one model allowing NSs up to $\sim 2 M_{\odot}$, while the competing model going up to $\sim2.5M_{\odot}$. When the injected mass is close to the maximum allowed mass and the SNR is sufficiently low, the posterior distribution is affected by this cutoff. In order to understand and visualize this effect, we construct a simple $1-$D model in Appendix~\ref{toymodel}. In the context of this simple model we explain both the effect on the BFs that favor the wrong model, and the BFs that decrease with increasing SNR. 

These results show that stacking many weak sources is \emph{not} necessarily equivalent to a single bright source. When each new observation is informative, i.e.~the likelihood dominates over the prior, then the data will prefer the correct model and BF $>1$. In that case, adding the extracted information from this observation will push the analysis in the right direction, and eventually, we will recover the same results as from a single loud event. On the other hand, if the observation is not informative and the result is prior dominated rather than likelihood dominated, the wrong model might be preferred and BF $<1$. In that case, adding this observation in the stack will lead the analysis in the wrong direction: a large number of weak observations that favor the wrong model will build confidence in the wrong conclusion.

The above results suggest that stacking must be treated with caution. For sufficiently high SNR events, the posterior will be narrow enough that it will not be affected by the maximum mass cutoff. In this case, one recovers the expected result: the correct model is preferred and it is preferred more and more as the signal strength increases. However, if the signals are of lower SNRs, the observations may not be informative, and then, the final cumulative result may be largely influenced by the prior and not by the new information contained in the signals.

\section{Comparing Equations of State}
\label{EOS}

\begin{table}
\begin{centering}
\begin{tabular}{c|c|cc}
\hline
\hline
\noalign{\smallskip}
Comparison & EoSs & $m_1(M_{\odot})$ & $m_2(M_{\odot})$ \\
\hline
\noalign{\smallskip}
 & SV/SV222 & 1.4 & 1.35\\
 &  & 1.95 & 1.9\\
kaons &  & 2.05 & 2\\ \cline{2-4}
 & GA-FSU2.1/GA-FSU2.1-180 & 1.9 & 1.8\\
 &  & 1.99 & 1.95\\
\noalign{\smallskip}
\hline
\noalign{\smallskip}
 & G4/H4 & 1.4 & 1.35\\
 &  & 1.8 & 1.7\\
hyperons &  & 1.95 & 1.9\\\cline{2-4}
 & MPa/MPaH & 1.95 & 1.9\\
 &  & 2.15 & 2.1\\
\noalign{\smallskip}
\hline
\noalign{\smallskip}
 & ALF-GCR/GCR & 1.5 & 1.4\\\cline{2-4}
 & ALF5/AP4 & 1.4 & 1.35\\
quarks &  & 1.8 & 1.7\\ \cline{2-4}
 & SQM3/AP4 & 1.2 & 1.1\\
 &  & 1.4 & 1.35\\
 &  & 1.8 & 1.7\\
\noalign{\smallskip}
\hline
\hline
\end{tabular}
\end{centering}
\caption{Simulated masses for the comparison of Sec.~\ref{EOS}. The first column gives the particle whose existence the comparison constrains, the second column gives the EoSs compared, while the third and fourth give the masses.}
\label{table:masses}
\end{table}

Our goal is to study whether GW detections of inspiraling NSs can be used to learn about their interior composition and in particular, whether they contain kaon condensates (Sec.~\ref{kaons}), hyperons (Sec.~\ref{hyperons}), and quark matter (Sec.~\ref{quarks}). To do so, we need to isolate their respective effects in the EoSs. We accomplish this by comparing pairs of EoS models that are as similar as possible, but differ only in the inclusion of one of these particles. Appendix~\ref{app-eos} provides a comprehensive classification of all the EoSs we use in this section.

\subsection{Kaon condensates}
\label{kaons}

We first address the question of whether NSs with kaon condensates in their inner cores leave an observable signature on inspiral GWs. To do so, we choose $3$ pairs of EoSs constructed with (i) the Skyrme-Hartree-Fock (SHF) scheme, (ii) relativistic mean field (RMF) theory, and (iii) the SHF scheme including three-nucleon interactions (TNI){\footnote{See Appendix~\ref{app-eos} for a short description of each method.}}. Each pair consists of one EoS with a kaon condensate and one without.

Kaon condensates could emerge in stars at high central densities and, therefore, kaon models differ from kaonless ones only for NSs with sufficiently high masses. This makes the extraction of physical information from these systems more difficult than that for low-mass systems for $3$ reasons. From a data analysis point of view, systems with masses close to the maximum allowed mass will suffer from the edge effects described in Sec.~\ref{stacking}, making it more difficult to get likelihood-dominated results. From a physical point of view, NSs with high masses have smaller values of $\bar{\lambda}$, as seen in Fig.~\ref{fig:love-m_all}, making finite-size effects less relevant in the GW phase. From an astrophysical point of view, NSs with masses around $2M_{\odot}$ are expected be rare. Table~\ref{table:masses} the values of the masses we select for each EoS comparison.

As we will show below, we find that if kaon condensates indeed form in the inner core of NSs, they will be hard to detect with the signals expected from aLIGO. On the other hand, if condensates do not form in NSs, then loud aLIGO signals may be able to establish this with a high mass detection.

\subsubsection{Models}
\label{kaonmodels}

\begin{figure}[t]
\begin{center}
\includegraphics[width=\columnwidth,clip=true]{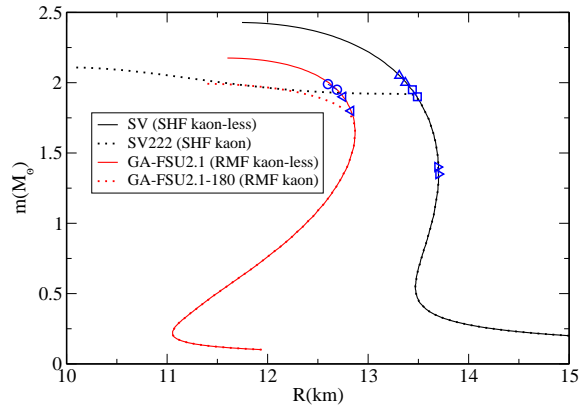}
\caption{\label{fig:mR_kaons}  m-R relation for the EoS pairs that test kaons: The black lines correspond to SV (solid) and SV222 (dotted), which are constructed through SHF models. The red lines are for GA-FSU2.1 (solid) and GA-FSU2.1-180 (dotted), which are constructed through RMF theory. The presence of kaons causes the kaon model of each pair to differ from the normal matter model for high masses. The pairs of similar blue symbols indicate the component masses we use in our analyses.}
\end{center}
\end{figure}
\begin{figure*}[t]
\begin{center}
\includegraphics[width=\columnwidth,clip=true]{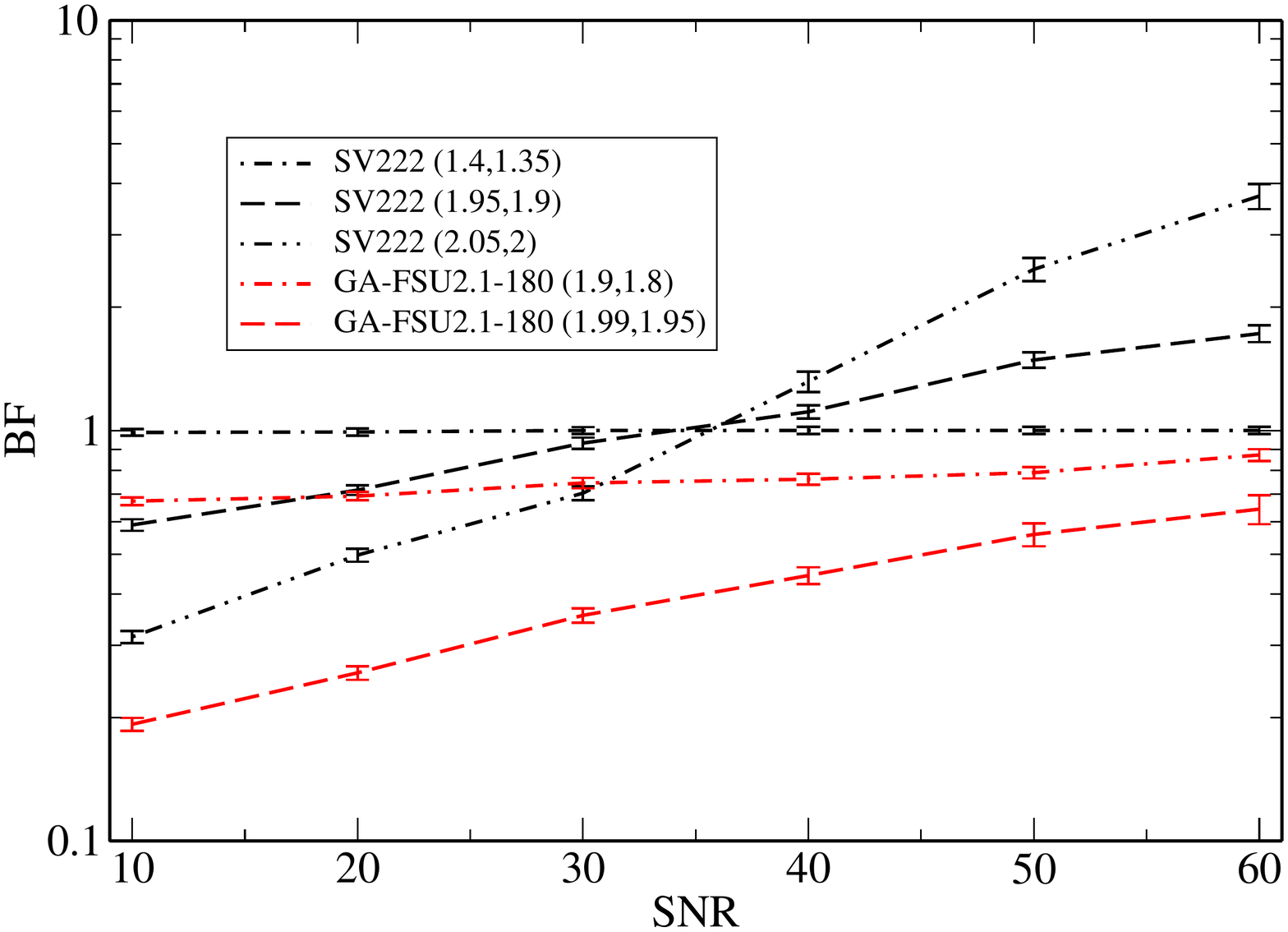}
\includegraphics[width=\columnwidth,clip=true]{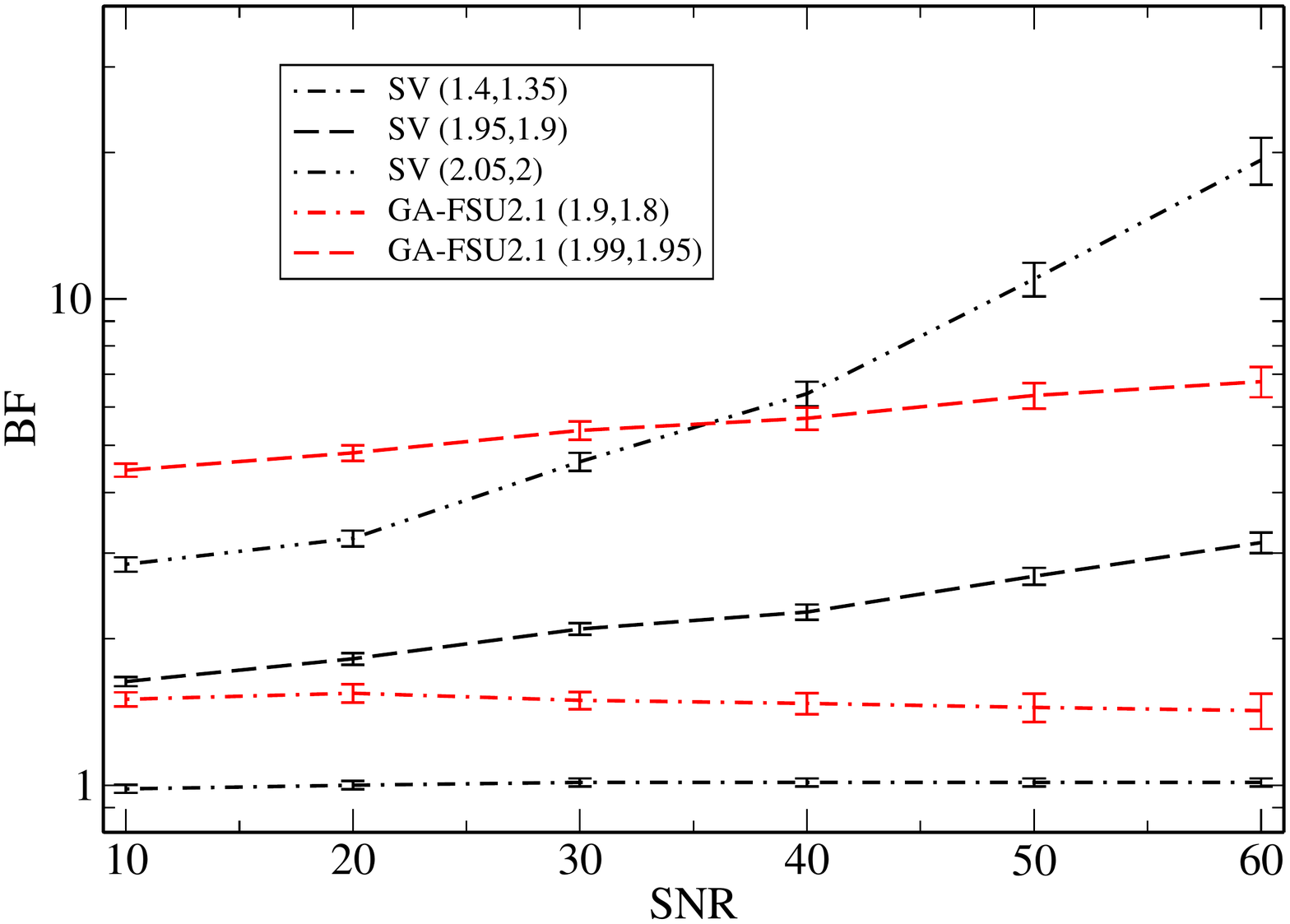}\\
\caption{\label{fig:bf_kaons}  BF in favor of the kaon models (left) and the kaonless models (right) as a function of the SNR for different injected masses given in the brackets. BFs are always quoted in favor of the correct (injected) model. For example, the black dotted-dashed curve labeled ``SV222 (1.4,1.35)'' in the left panel means that the injected model is the SV222 EoS with the NS mass $(1.4,1.35)M_\odot$, and $\mathcal{H}_1 = \mathrm{SV222}$ and $\mathcal{H}_2 = \mathrm{SV}$ in Eq.~\eqref{bf}. BF $> 1$ in the left panel means that we correctly detected the presence of kaons inside the NSs, while that in the right panel means that we correctly ruled out such a presence of kaons. The kaonless models give higher BFs than the kaon models, making it easier to establish that NSs do not have kaon condensates than the opposite. BF $<1$ in the low SNR regime for high-mass systems in the left panel is due to the fact that the injected masses are close to the maximum mass allowed in each model (see the main text for more detail).}
\end{center}
\end{figure*}

The $3$ pairs of models we compare are constructed with different approximations. 
\begin{enumerate}
\item \emph{SHF models}~\cite{Lim:2013tqa}: 
\begin{enumerate}
\item kaonless: SV.
\item with kaons: SV222. 
\end{enumerate}
\item \emph{RMF theory}~\cite{2013JPhG...40b5203G}: 
\begin{enumerate}
\item kaonless: GA-FSU2.1. 
\item with kaons GA-FSU2.1-180.
\end{enumerate}
\item \emph{SHF models + TNI}~\cite{Lim:2013tqa}: 
\begin{enumerate}
\item kaonless: SGI. 
\item with kaons: SGI178. 
\end{enumerate}
\end{enumerate}
The results of comparison (3) are very similar to the results of comparison (2), so we will only present pairs (1) and (2). Figure~\ref{fig:mR_kaons} shows the mass-radius relation for the EoSs we present. This figure suggests that kaons affect the properties of only the most massive NSs. The pairs of circles, triangles, and squares indicate the values of the masses in the injected signals of our analysis. 

\subsubsection{Bayes factor}
\label{kaonbf}

Figure~\ref{fig:bf_kaons} shows the BF in favor of the kaon model of each pair (left panel) and the kaonless model of each pair (right panel) as a function of the SNR of the signal. The different lines correspond to the different injected masses indicated by the blue symbols in Fig.~\ref{fig:mR_kaons}. We always plot the BF in favor of the correct (injected) model. Any BF $> 1$ in the left panel means that we correctly detected the presence of kaons in the interior of the NSs, while any BF $> 1$ in the right panel means that we correctly concluded that there are no kaons in their interiors. 

As first suggested by the mass-radius plot, Fig.~\ref{fig:bf_kaons} confirms that in order to detect the presence of kaons on NS EoS we need a high mass system. The lowest mass system $(m_1=1.4M_{\odot},m_2=1.35M_{\odot})$ gives BFs that are consistent with $1$, in agreement with the m$-$R relation. For large SNRs, as we increase the injected masses we recover BFs that correctly favor the kaon model (left panel) and correctly favor the kaonless models (right panel). 

At low SNRs, however, we encounter BFs that incorrectly disfavor the kaon models on the left panel of Fig.~\ref{fig:bf_kaons}. This is because the injected masses are very close to the maximum mass allowed by the correct model. When an injected parameter is closer to the edge of the prior than the width of the posterior, the posterior has to be cut off (see Sec.~\ref{stacking} and Appendices~\ref{cutoffs} and~\ref{toymodel}). In our case, the injected masses are close enough to the maximum mass of the kaon models that these models ``lose" some chain points because the models simply cannot produce such high mass systems. What the kaon models are in fact trying to do when the mass is above their maximum allowed mass is match the GW signal with pure Gaussian noise. How this affects the BF is clear from Eq.~\eqref{bf-def}. Recall that the BF in favor of SV222, for example, is equal to the number of times the chains visited this model, divided by the number of times they visited the competing model SV. When SV222 has an abrupt cutoff close to the injected masses, a lot of the chain points will be disfavored because the model fails to produce a signal for these values of the masses. This leads to the chains visiting the SV222 model less, and in the end a BF that does not favor the correct model. 

The discussion of Sec.~\ref{stacking} indicates that when we have BFs favoring the wrong model, we are in a regime where it is mostly the prior that dominates our results. Obviously, the results in this regime do not offer new physical information and cannot be used to claim that we constrained the EoS. For example, imagine we detected a $(1.99,1.95)M_{\odot}$ system at SNR 30 and we want to claim something about the presence or absence of kaons in the system. If we choose to test whether GA-FSU2.1 or GA-FSU2.1-180 fit the data better, we will recover BFs in favor of the kaonless model GA-FSU2.1 regardless of whether kaons are actually present or not. The data coming from such a system can clearly not be trusted to give the correct result.

For the case studied here, the kaon model comparison starts to become likelihood-dominated (BF $> 1$) when the SNR $\gtrsim 40$ in the SV/SV222 case, and when the SNR is somewhere above $60$ in the GA-FSU2.1/GA-FSU2.1-180 case. A detection of a $(2.05,2)M_{\odot}$ system at SNR $= 40$ and above will lead to a strong BF in the Jeffreys scale in favor of the kaonless model for a kaonless injection. On the other hand, for a kaon injection, the SNR needs to reach 60 and above in order to obtain strong BFs. For lower mass systems, the BFs are lower, and barely worth mentioning in the Jeffreys scale.

\subsection{Hyperons}
\label{hyperons}

Moving on to the study of the detectability of hyperons in the inner cores of NSs, we select $5$ EoSs constructed with (i) RMF theory, (ii) a nonrelativistic Brueckner-Hartree-Fock (BHF) approach, (iii) a relativistic BHF approach, (iv) and (v) a SHF approach. Each pair consists of one EoS with a hyperon in the inner core and one EoS without.

Hyperons, much like kaons, affect the EoS of only the most massive NSs, since they form at the very highest central densities. For that reason, the results of this section are very similar to the previous one on kaons: we find that aLIGO can constrain the existence of hyperons in NSs, but detecting them will be much harder (see Table~\ref{table:masses}).

\subsubsection{Models}
\label{hyperonmodels}

\begin{figure}[t]
\begin{center}
\includegraphics[width=\columnwidth,clip=true]{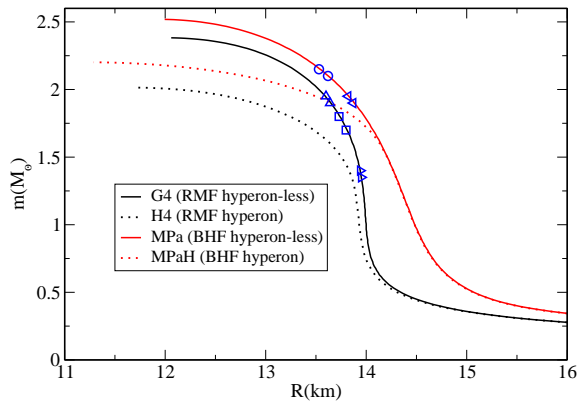}
\caption{\label{fig:mR_hyperons}  m-R relation for the EoS pairs that test hyperons: The black lines correspond to G4 (solid) and H4 (dotted), which are constructed through RMF theory. The red lines are for Mpa (solid) and MPaH (dotted), which are constructed through the nonrelativistic BHF method. The presence of hyperons causes the hyperon model of each pair to differ from the normal matter model for high masses. The pairs of similar symbols indicate the component masses we use in our analyses.}
\end{center}
\end{figure}

The models we use to determine hyperon detectability are the following.
\begin{enumerate}
\item \emph{RMF theory}~\cite{Lackey:2005tk}: 
\begin{enumerate}
\item hyperonless: G4. 
\item with hyperons: H4.
\end{enumerate}
\item \emph{Nonrelativistic BHF models}~\cite{2014PhRvC..90d5805Y}: 
\begin{enumerate}
\item hyperonless: MPa, 
\item with hyperons: MPaH.
\end{enumerate}
\item \emph{relativistic BHF models}~\cite{2014arXiv1410.7166K}: 
\begin{enumerate}
\item hyperonless: DBHF$^{(2)}$(A).
\item with hyperons: NlY5KK$^{*}$.
\end{enumerate}
\item \emph{SHF model}: 
\begin{enumerate}
\item hyperonless: SGI~\cite{Lim:2013tqa}, 
\item with hyperons: SGI-YBZ6-S$\Lambda$$\Lambda$3~\cite{2014arXiv1412.5722L}. 
\end{enumerate}
\item \emph{SHF models}: 
\begin{enumerate}
\item hyperonless: SkI4~\cite{1995NuPhA.584..467R}, 
\item with hyperons: SkI4-YBZ6-S$\Lambda$$\Lambda$3~\cite{2014arXiv1412.5722L}. 
\end{enumerate}
\end{enumerate}
The results of comparisons (3), (4) and (5) are very similar to the results of the MPa and MPaH comparison, so we will not present them here. Figure~\ref{fig:mR_hyperons} shows the mass-radius relation for the EoSs for which we will present comparisons. Clearly, hyperons affect the EoS of only the most massive NSs, exactly like kaons. The pairs of symbols indicate the masses we inject. 

\subsubsection{Bayes factors}
\label{hyperonbf}
\begin{figure*}[t]
\begin{center}
\includegraphics[width=\columnwidth,clip=true]{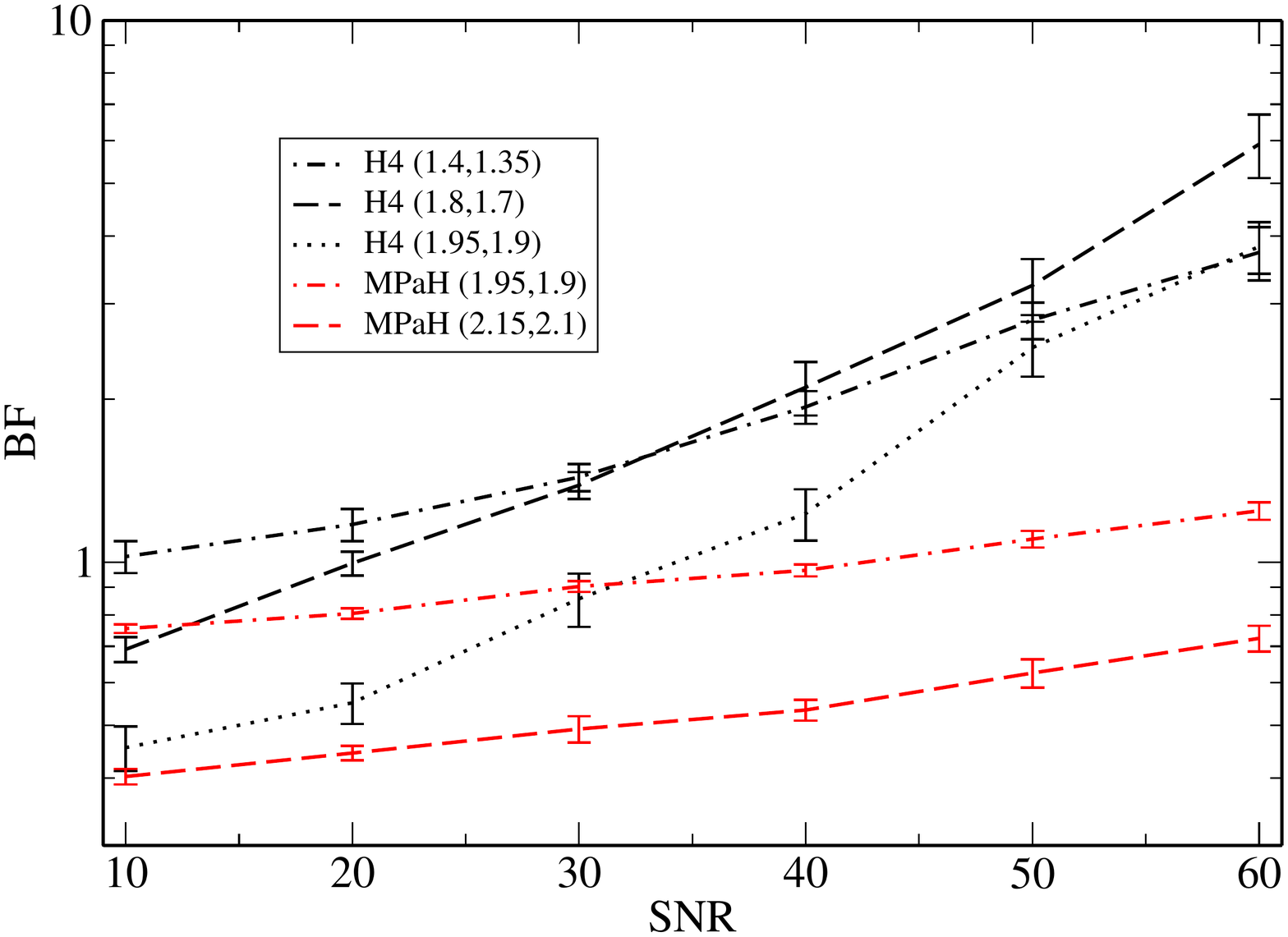}
\includegraphics[width=\columnwidth,clip=true]{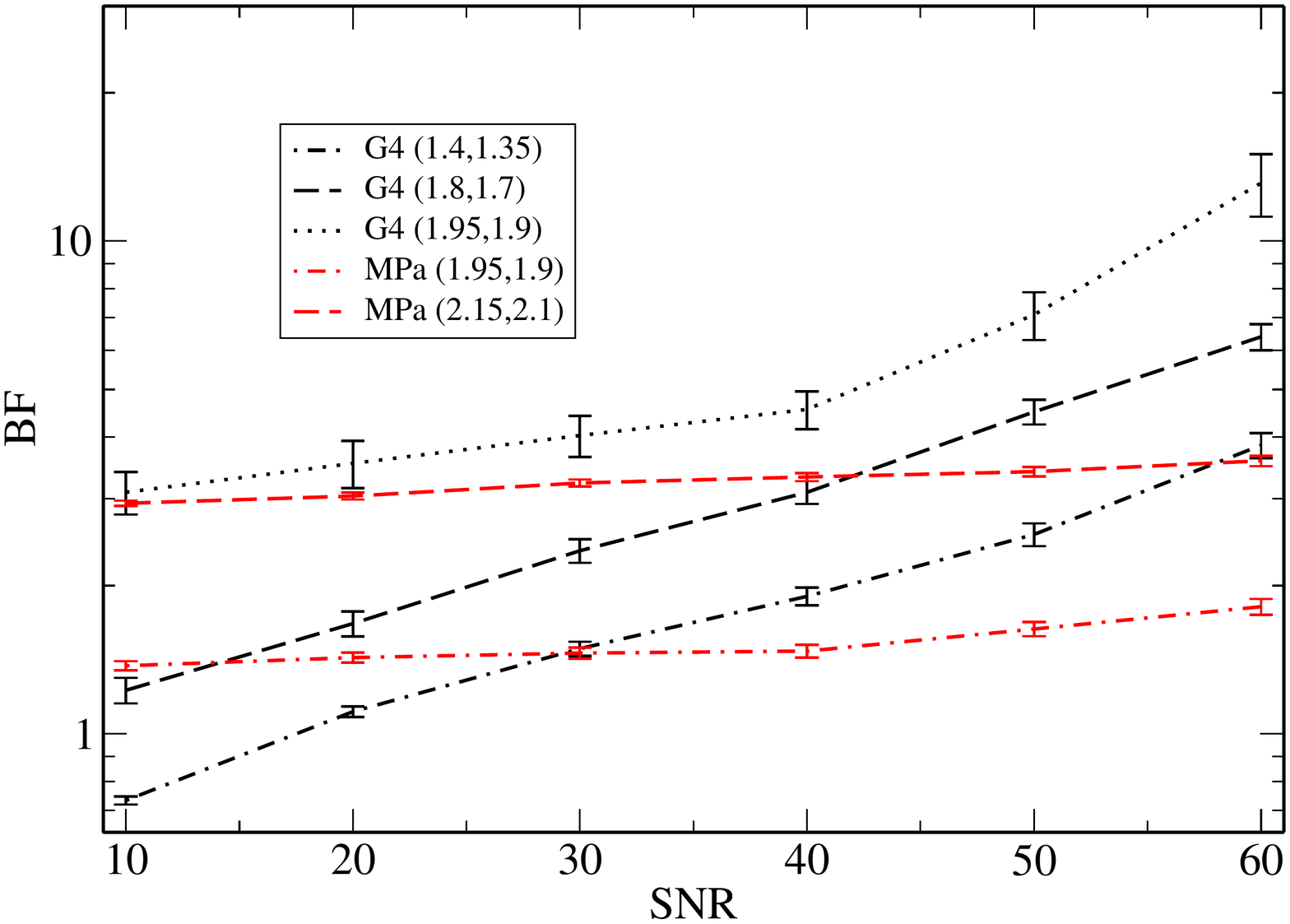}\\
\caption{\label{fig:bf_hyperons}  BF in favor of the hyperon models (Left) and the hyperonless models (Right) as a function of the SNR for different injected masses. BFs are always quoted in favor of the correct (injected) model. We conclude that it is easier to constrain hyperons than detect them. A similar result was reached in the case of kaons as shown in Fig.~\ref{fig:bf_kaons}.}
\end{center}
\end{figure*}

The results of our analysis are presented in Fig.~\ref{fig:bf_hyperons}, which shows the BF in favor of the hyperon models (left) and the hyperonless models (right) as a function of the SNR of the injected signal. 
The message of this plot is clear if we take into account the reasoning presented in the previous section. In the G4-H4 comparison for the hyperonless comparison at low SNR (below 40), we encounter the effect of the wrong model being preferred over the correct one. On the other hand, in the MPa-MPaH comparison, this continues to hold until the SNR $=60$. This indicates that it is the prior (and more specifically the maximum mass of the hyperon model) that dominates our results and not the likelihood. When the SNR exceeds 40, we start extracting interesting information from the comparison. Assuming hyperons do not form in NS cores, a $(1.95,1.9)M_{\odot}$ detection with SNR $=40$ and $60$ will give a strong and very strong indication in the Jeffreys classification of the nonexistence of hyperons respectively. On the other hand, if hyperons do form in NS cores, then a signal with SNR $=60$ would only provide \emph{strong} evidence, as the BFs do not exceed 7. 

\subsection{Quark matter}
\label{quarks}

Unlike kaon condensates and hyperons that can only exist in combination with normal matter, quark matter can form both with and without normal matter. In the first case we have hybrid NSs with EoSs of the ALF~\cite{Alford:2004pf} type that have quarks formed after a certain transition density. The second case results in quark stars~\cite{SQM} comprised solely of quark matter. 

The EoSs of pure quark stars differ from normal matter EoSs even at low densities, making it possible for aLIGO to detect or rule them out. On the other hand, hybrid normal/quark matter EoSs are constructed by stitching the nuclear matter EoSs in the low density regime to quark matter ones in the high density regime, with appropriate phase transitions in between. Therefore, they reduce to normal matter EoSs at low densities. The transition density $n_c$ and the strength of the strong interactions $c$ determine how much the hybrid EoS differs from the normal matter EoS it is stitched to. The constant $c$ is predicted to be $c \sim 0.37$~\cite{Fraga:2001id} by a perturbative calculation. When $c$ is close to this value and the transition from a nuclear matter to quark matter happens at around twice the saturation density, we find that hybrid EoSs might be distinguishable from normal matter EoSs for signals with SNR $\sim 30-40$. 

\subsubsection{Hybrid EoSs}

\begin{figure*}[t]
\begin{center}
\includegraphics[width=\columnwidth,clip=true]{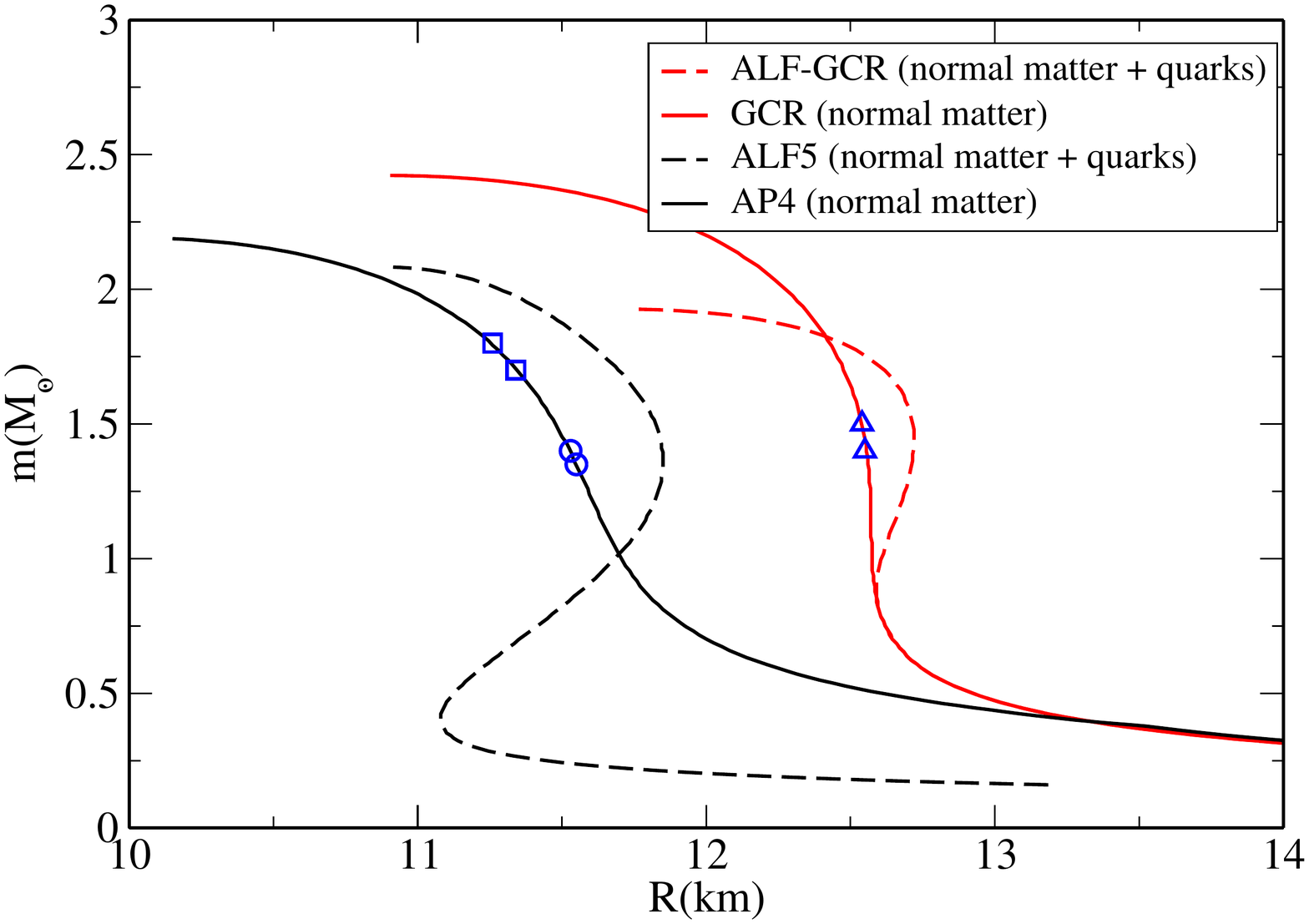}
\includegraphics[width=\columnwidth,clip=true]{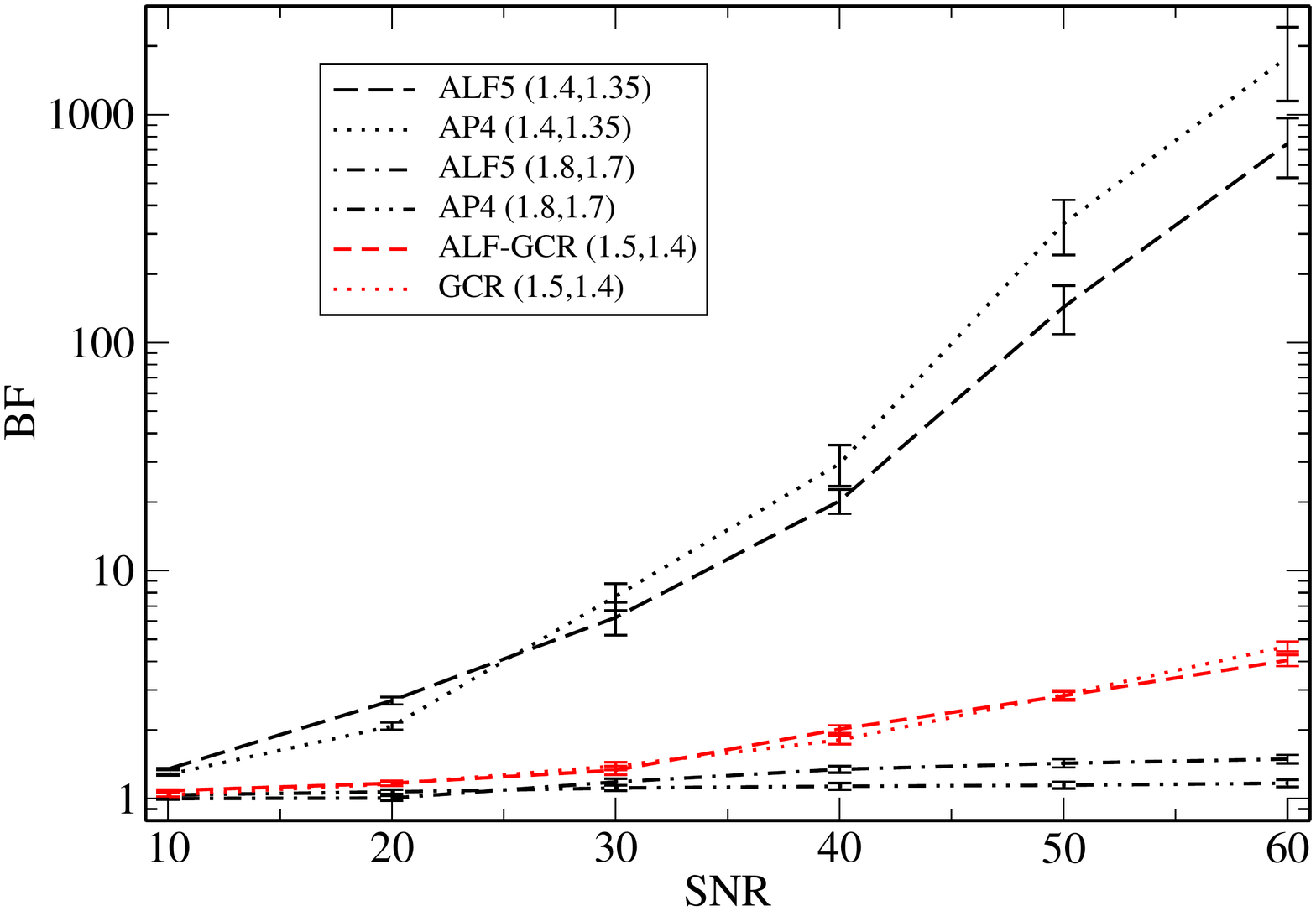}
\caption{\label{fig:ALF2AP4}  (Left) m-R relation for the hybrid EoSs and the nuclear matter EoSs they are based on and compared to: AP4 (solid black), ALF5 (dashed black), GCR (solid red), and ALF-GCR (dashed red). The pairs of similar symbols denote the injected masses. The larger the value of the strong interactions, the larger the deviation between the normal matter and the hybrid EoS. (Right) BF for the ALF5/AP4 comparison (black lines) and the ALF-GCR/GCR comparison (red lines) as a function of the SNR for different injected masses. BFs are given in favor of the correct model denoted in the legend. When the strong interactions between the quarks are close to the value predicted from perturbative calculations, it is possible to distinguish between normal and hybrid NSs.}
\end{center}
\end{figure*}

In order to study hybrid normal/quark matter stars we select EoSs of the ALF family~\cite{Alford:2004pf}, the normal matter part of which are based on AP4~\cite{APR}. All EoSs include ordinary matter, while the ALF EoSs also include pions and quarks in the inner core\footnote{In principle, we should treat pions as a separate particle, like kaons and hyperons. However, we are not aware of any EoS model that includes only pion condensates and predicts a maximum NS mass above $2M_{\odot}$.}. The left panel of Fig.~\ref{fig:ALF2AP4} shows the mass--radius relation for ALF5 and AP4. ALF5 has $(n_c,c)=(2n_0,0.4)$, where $n_0 = 0.16 \text{ fm}^{-3}$ is the nuclear saturation density. 

We also constructed new hybrid star EoSs (GCR-ALF) by stitching the nuclear matter GCR EoS constructed in~\cite{Gandolfi:2011xu} to the same quark matter EoS as the ALF family. The m-R relation for a GCR-ALF EoS with $(n_c,c) = (2 n_0,0.35)$ is shown again in Fig.~\ref{fig:ALF2AP4}, together with the corresponding nuclear matter GCR EoS with the symmetry energy of $E_\mathrm{sym}=33.8$MeV. We have also chosen different values for $n_0$, $c$ and $E_\mathrm{sym}$ and found that the difference between the nuclear matter and hybrid EoSs are typically even smaller than the one in Fig.~\ref{fig:ALF2AP4}.

 The right panel of Fig.~\ref{fig:ALF2AP4} shows the BFs we recover for the ALF5/AP4 (black) and the ALF-GCR/GCR comparison (black) as a function of the SNR for different values of the masses. The only case we find where the hybrid EoS could be distinguishable from the normal matter one is for masses around $1.4M_{\odot}$. A SNR $\sim30$ detection with such masses can provide strong BFs in the Jeffreys scale if we compare ALF5 to AP4. However, if we compare ALF-GCR to GCR we recover smaller BFs. We also find that we recover similar results when comparing hybrid stars and a normal NS regardless of which one is the correct star of Nature. This is different from the kaon and hyperon cases studied before, where we obtain more conclusive BFs when kaons or hyperons are not present in NSs. Of course, if the true hybrid EoS of Nature contains weaker strong interactions between the quarks, the prospects of detecting a hybrid star reduce even further.

\subsubsection{Quark stars}
\begin{figure*}[t]
\begin{center}
\includegraphics[width=\columnwidth,clip=true]{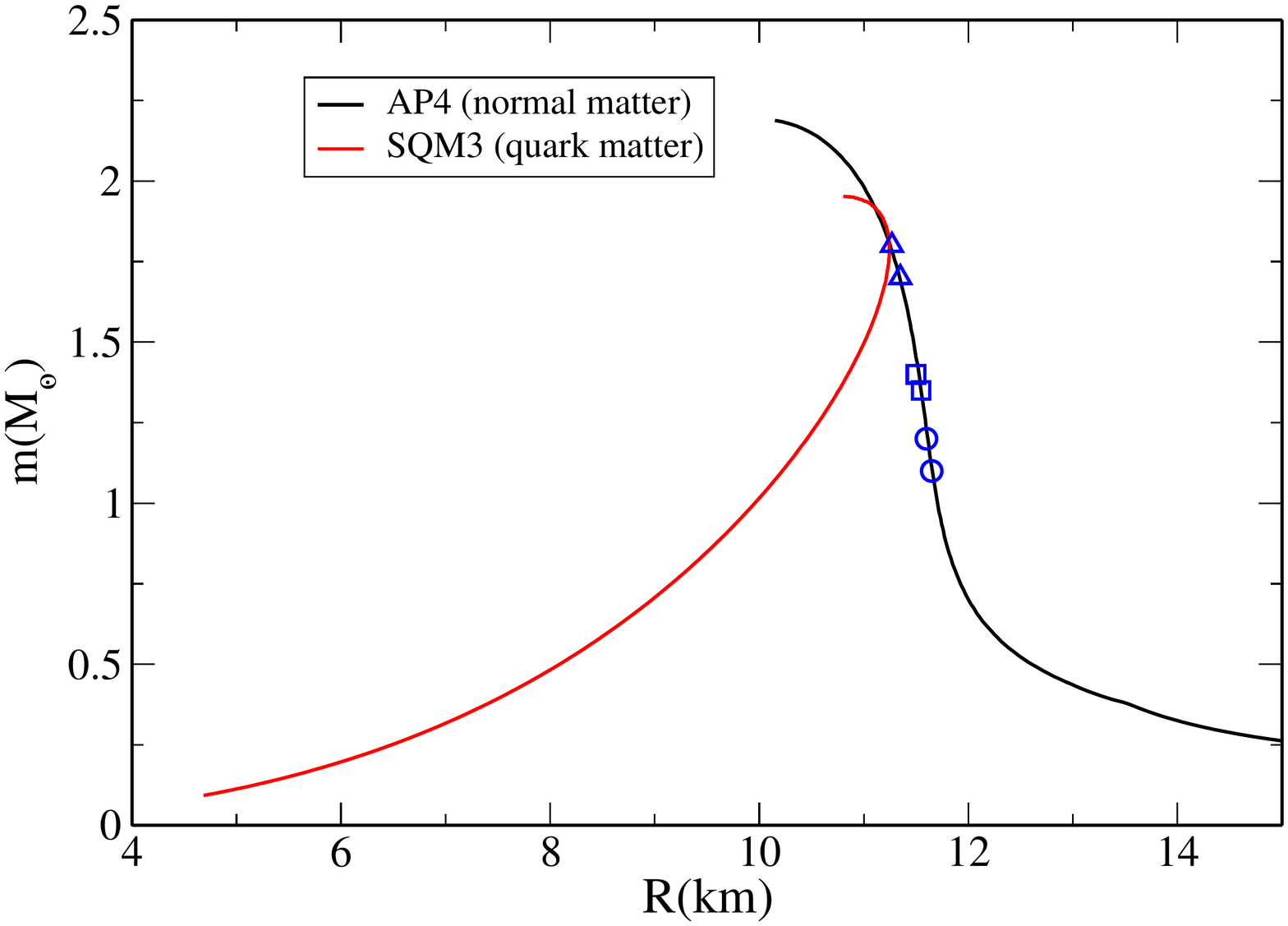}
\includegraphics[width=\columnwidth,clip=true]{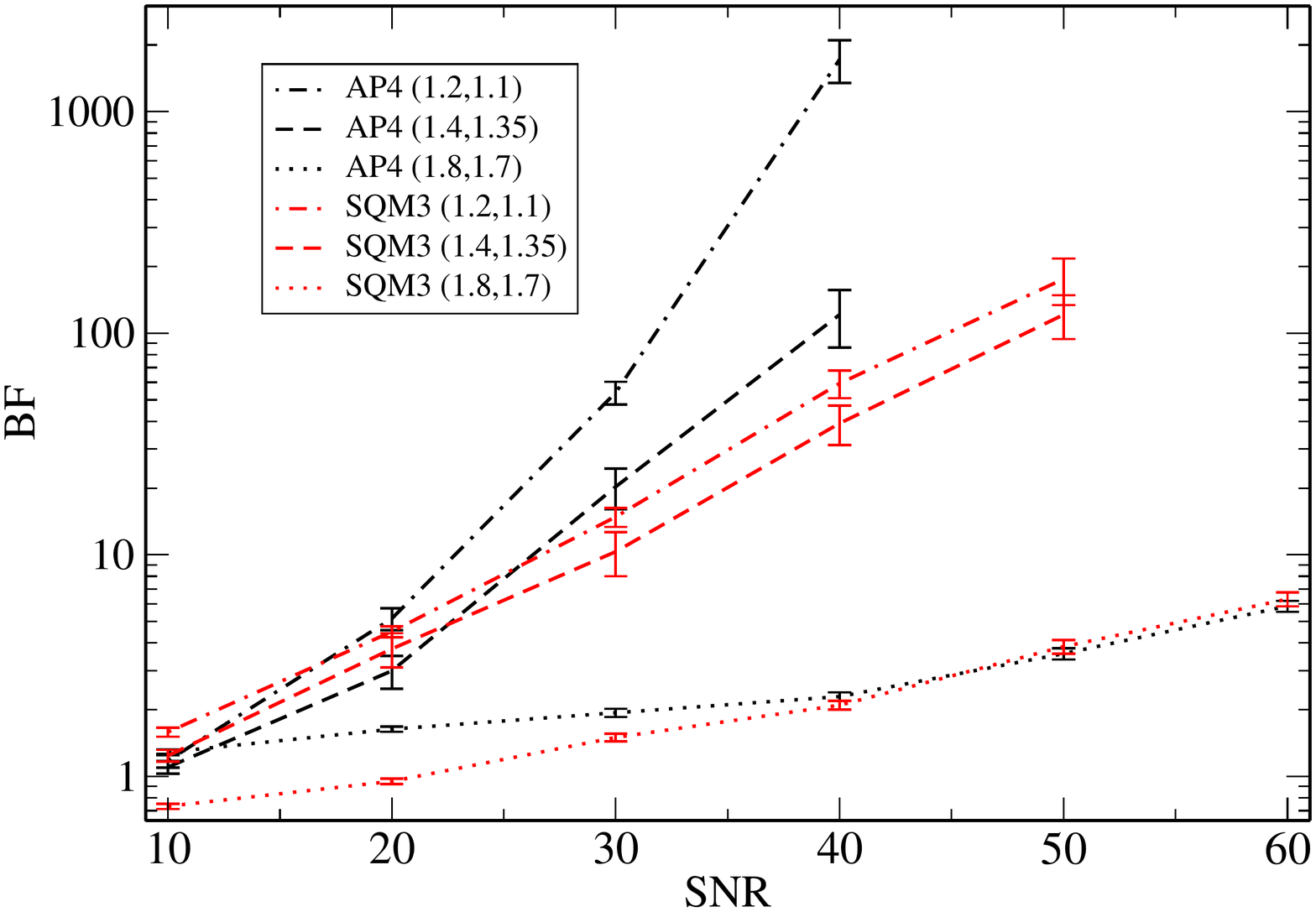}
\caption{\label{fig:SQM3AP4}  (Left) m-R relation for SQM3 and AP4. The pairs of similar symbols denote the injected masses. (Right) BF in favor of AP4 (black) and SQM3 (red) as a function of the SNR for different injected masses. BFs are given in favor of the correct model. aLIGO will be able to place strong constraints on the existence of strange quark stars, both detecting them if present in Nature or strongly disfavoring their existence if not.}
\end{center}
\end{figure*}

Our final study case is SQM3~\cite{SQM}, an EoS that contains no normal matter at all, but rather it is constructed solely with quark matter. Comparing it to a normal matter EoS amounts to comparing normal NSs to strange quark stars. Although the latter have already been heavily constrained~\cite{Alford:2015dpa}, it is still interesting to study the bounds aLIGO could place on them. However, it is not clear what EoS we should compare SQM3 to. Throughout our analysis, we compare EoSs that contain exotic matter to the EoS that we obtain if we remove the exotic matter, but change nothing else in how it is constructed; this is precisely how we defined EoS pairs. In the case of SQM3, if we remove the quarks there is no matter left, so there is no normal matter counterpart that can form a pair with a strange quark star.

For that reason and in order to arrive at conservative conclusions, we will compare SQM3 to AP4~\cite{APR}, a reliable EoS that has both not yet been ruled out by observations and leads to NSs that are the most similar to SQM3 quark stars in the set of EoSs we considered. The left panel of Fig.~\ref{fig:SQM3AP4} shows the mass--radius relation for SQM3 and AP4, along with symbols that indicate injected masses. Being the softest EoS in the set of nonruled out EoSs we considered, the AP4 mass-radius relation is the farthest to the left in Fig.~\ref{fig:SQM3AP4} and thus the closest to the SQM3 one.  Other normal matter EoSs are stiffer than AP4, predicting larger radii for the same mass, which leads to m $-$ R curves even farther away from SQM3 than AP4. As we show below, AP4 is already clearly distinguishable from SQM3, and thus, distinguishing between SQM3 and other stiffer normal-matter EoSs would be even easier. 

Clearly, SQM3 is very different from all other EoSs studied here; its m $-$ R relation differs qualitatively from normal matter EoSs, and we thus expect it to be distinguishable. 
Observe that SQM3 fails to produce a $2M_{\odot}$ NS, though it is still consistent with the current observational bound on the maximum NS mass~\cite{2.01NS} within a 2-$\sigma$ statistical error. 
Of course, it is possible that the star observed in~\cite{2.01NS} is simply a NS, and not a quark star, without necessarily ruling out the existence of the latter. We have thus decided to study whether aLIGO can distinguish between strange quark stars and normal NSs. If all compact stars that aLIGO sees are NSs or BHs, then this would build confidence for the nonexistence of quark stars.

The right panel of Fig.~\ref{fig:SQM3AP4} shows the BF in favor of AP4 (black) and SQM3 (red) as a function of the SNR of the signal for different mass combinations. The really high BFs we recover indicate that aLIGO will be able to both detect strange quark stars if they exist or provide very strong evidence for their nonexistence. For example, the detection of a $(1.4,1.35)M_{\odot}$ system with SNR $=20$ results in strong BFs in the Jeffreys scale in favor of the correct model. For even brighter sources, or less massive systems, we recover very strong or even decisive evidence in favor of the correct model of Nature, be it strange quark stars or normal NSs. 

We have argued that the results given above represent the worst case scenario when comparing quarks stars to normal NSs, given the reliable normal matter EoSs available today and not yet ruled out by observations. But what if nuclear physicists construct a viable EoS that is softer than AP4? Even in this scenario, we can make some claims based on the unique shape of SQM3. Revisit the left panel of Fig.~\ref{fig:SQM3AP4} and notice that between $1.8M_{\odot}$ and $1M_{\odot}$ the radius SQM3 predicts increases by about $2$km. A normal matter EoS that is softer than AP4 will, roughly speaking, have a similar shape to AP4, but it will be shifted to the left in the m $-$ R plane. Even in this case, there will exist some mass between $1M_{\odot}$ and $1.8M_{\odot}$ where the difference between the radius of a quark star and a normal NS is at least $1$km. Our results indicate that typically a radius difference around $0.5$km is about enough for distinguishability for this type of study, provided the mass is not close to the maximum mass allowed. We can, therefore, claim that even if a softer than AP4 EoS is constructed, it will still be distinguishable from SQM3 for some value of the mass between $1$ and $1.8 M_{\odot}$, given an SNR around 30, depending on its exact shape.

\subsection{Edge effects}

So far, we have injected signals with masses for which both models can produce a NS (or a hybrid star or a quark star). In the case of hyperons and kaons, we saw that the large value of the masses required to tell EoS models apart are close to the maximum mass of the exotic matter model. Low strength signals are greatly affected by this maximum mass cutoff. Here, we examine the case where one of the two objects can only be supported by the model that predicts the higher maximum mass.
\begin{figure}[t]
\begin{center}
\includegraphics[width=\columnwidth,clip=true]{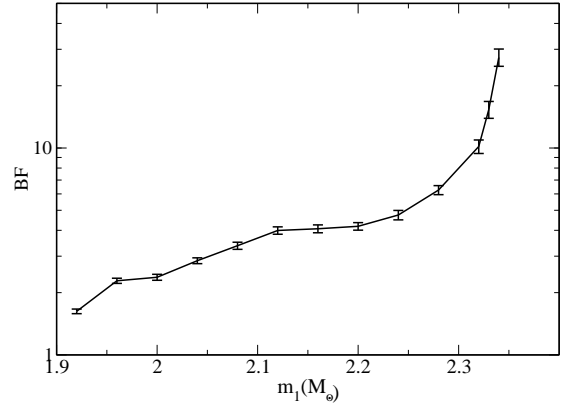}
\caption{\label{fig:BFSV0m214}  BF in favor of SV (a kaonless EoS) compared to SV222 (a kaon EoS) as a function of $m_1$ for $m_2=1.9 M_\odot$ with SNR$=30$.}
\end{center}
\end{figure}
\begin{figure}[t]
\begin{center}
\includegraphics[width=\columnwidth,clip=true]{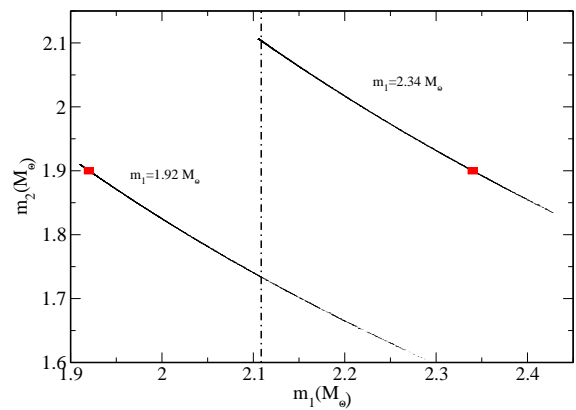}
\caption{\label{fig:m1m2}  $2$D Scatter plot in the $m_1-m_2$ plane for injected masses $m_2=1.9M_{\odot}$ and $m_1=1.92M_{\odot}$ and $m_1=2.34M_{\odot}$. The red box indicates the injected masses and the vertical line gives the maximum mass SV222 can support.}
\end{center}
\end{figure}

We revisit the pair SV-SV222 and fix the SNR to 30 and the mass of the smaller NS in the binary to $1.9 M_{\odot}$. We inject signals constructed with SV and in Fig.~\ref{fig:BFSV0m214} we plot BF in favor of the correct model as a function of the NS mass of the larger star $m_1$. The maximum mass supported by SV222 is $\sim 2.11 M_{\odot}$. When both simulated masses can be supported by the wrong model, then we find BFs that are only barely worth mentioning in the Jeffreys scale in favor of the correct model. However, when $m_1 >2.11 M_{\odot}$ the BF in favor of SV starts increasing and it diverges around $m_1=2.35 M_{\odot}$. At this point, the chirp mass of the system is so large that it cannot be matched by a system with both masses below $2.11 M_{\odot}$. Beyond this point, SV222 cannot produce any systems with the correct chirp mass of the injected signal.

To illustrate this transition we plot the $2-$D scatter plot of the chain points in the $m_1-m_2$ plane in Fig.~\ref{fig:m1m2} for $m_1=1.92M_{\odot}$ and $m_1=2.34M_{\odot}$. 
 The red boxes mark the injected masses and the vertical line is the maximum mass SV222 can support. The scatter plots have support only along constant chirp mass lines. For $m_1=1.92M_{\odot}$ most points fall on the left of the vertical line, and SV222 can provide a good match for them. However, as we increase $m_1$, more and more points move beyond SV222's maximum mass, resulting in BFs that favor it less and less (recall that the BF, as given in Eq.~\eqref{bf-def}, is the ratio of the points in SV222 over the points in SV). When $m_1=2.34M_{\odot}$ only a small number of points can be supported by SV222, and we recover very strong BFs in favor of SV. If we increase the mass even more, no points fall on the left of the vertical line, and the BF goes to infinity. 

\subsection{Noise curves}
\label{noise}

\begin{figure}[t]
\begin{center}
\includegraphics[width=\columnwidth,clip=true]{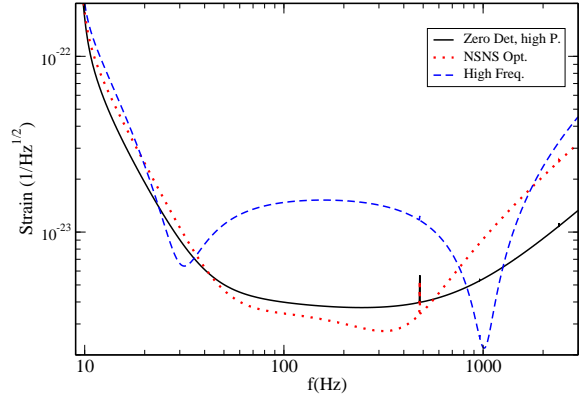}
\caption{\label{fig:noisecurve}  Sensitivity curves of various aLIGO configurations. The High-Freq. optimized curve (blue dashed line) is very sensitive in a small window around $1000$ Hz, but it has much higher noise at lower frequencies. The NSNS Opt. noise curve (red dotted line) has slightly lower noise at frequencies below $600$ Hz, but much higher noise above this. EoS effects become important at frequencies above $400$ Hz.}
\end{center}
\end{figure}
\begin{figure}[t]
\begin{center}
\includegraphics[width=\columnwidth,clip=true]{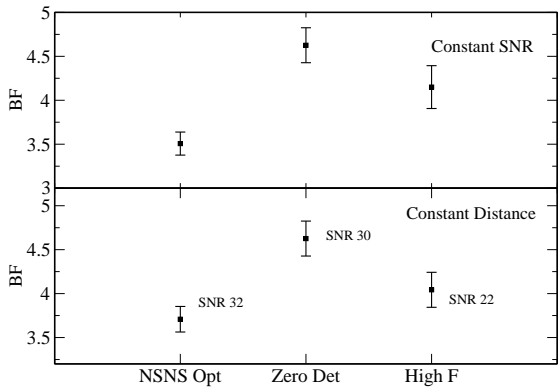}
\caption{\label{fig:bfnoise}  BF for the same system with the different aLIGO noise curves. In the top panel we keep the SNR constant, while at the bottom panel we keep the luminosity distance constant. The SNR to which this distance corresponds with each noise curve is indicated in the plot. In both cases, the Zero-Det., High-P.~configuration gives the highest BFs at these distances and SNRs, making it the optimal noise curve for EoS studies.}
\end{center}
\end{figure}

Apart from the Zero-Det., High-P.~noise configuration, there are a few 
other tuned noise curves for aLIGO~\cite{AdvLIGO-noise}. Among them, the NSNS 
Optimized configuration gives the optimal SNR for a NSNS coalescence, while the 
high frequency one achieves the best sensitivity around 1000Hz. 
Figure~\ref{fig:noisecurve} shows these $3$ noise curves. The NSNS Opt.~noise curve is tuned to 
give the highest SNR by reducing the noise levels around $(60-500)$Hz, however 
this comes at the expense of higher noise in the kHz region. The finite-size 
effects that we are looking for manifest themselves at frequencies above $400$ 
Hz, suggesting that NSNS Opt. is suboptimal for EoS determination. On the other 
hand, the High F.~configuration has the lowest noise in a small window around $10^3$Hz, but it is 
clearly inferior for lower frequencies. This loss of SNR at low frequencies 
makes its suitability for EoS studies questionable.

Figure~\ref{fig:bfnoise} demonstrates how the standard High-P., 
Zero-Det.~configuration is the optimal noise curve for measuring the EoS both 
for systems of constant SNR (top panel) and for systems at the same distance 
(bottom panel). We plot the BF for a system with the same parameters (apart from 
the distance in the top panel) achieved with 
the $3$ noise configurations. In the constant distance case, NSNS Opt. achieves 
the best SNR value while High F. has the worst, as expected. In both cases, the 
lower high-frequency noise of the Zero-Det., High-P.~configuration together with 
its low-enough noise at lower frequencies outperforms both tuned configurations 
in EoS extraction. This is another manifestation of the effect discussed 
in~\cite{Sampson:2014qqa}; when we are looking for an effect that appears only 
at certain frequencies, it is not the total SNR that matters, but the SNR 
accumulated in those frequencies.

\section{Conclusions}
\label{discussion}

We studied whether future GW data from the advanced, ground-based detectors will allow us to learn about the internal composition of NSs. We find that aLIGO can efficiently distinguish between NSs with EoSs that differ at relatively low to moderate central densities. This is the case, for example, for EoSs that model just pure quark matter. If the NS is a hybrid of normal and quark matter, higher SNR values are required, as well as relatively strong interactions between the quarks. On the other hand, aLIGO will not be able to efficiently distinguish between NSs with EoS that differ only at high central densities. This is the case, for example, for EoSs that model normal matter and those that include a hyperon or a kaon condensate in the inner core. In this case, aLIGO would require a very loud detection to be able to discern between normal and exotic matter NSs. 

In this paper, we considered kaon, hyperon, and hybrid EoSs with exotic matter parameters within the range allowed from experiments and theoretical calculations, that show relatively large difference from the corresponding normal matter EoS. However, these exotic matter parameters can be varied within their experimental range to construct exotic matter EoSs that are more similar to their corresponding normal matter EoS. If one were to marginalize the BFs over such exotic matter parameters by considering as many exotic matter EoSs as possible with different choices of parameters, calculating the BFs and taking the average, the main results in this paper would still hold. In fact, such an analysis would strengthen our conclusion since one would find that it would be more difficult for aLIGO to distinguish between normal and exotic matter EoSs than the results presented in the main text. 

Our results could be improved if an accurate description of the merger, where the finite size effects are more prominent, is included in the models. For example, the effect of hyperons and strange quark matter on GWs from merging NS binaries have been studied e.g. in~\cite{Hotokezaka:2011dh} and~\cite{Bauswein:2009im} respectively. However, NS mergers always occur at kHZ frequencies\footnote{The NS merger frequency cannot be pushed arbitrarily low since the NS mass cannot exceed roughly $3 M_{\odot}$ by causality.}, where the detector sensitivity is lower and it is the inspiral phase that falls in the most sensitive frequency band. Moreover, to this day, no complete template bank of NSNS mergers exists. It would be interesting to study how much information can be extracted from the merger phase, and how this could aid EoS determination by carrying a full inspiral-merger analysis.

As a side note, we find that the aLIGO noise configuration that maximizes the gain of relevant physical information about NS EoS is the standard Zero-Detuned, High-Power one. This is perhaps contrary to the belief that a high-frequency tuned noise configuration would do best. The reason why the Zero-Detuned, High-Power configuration does better is that the finite-size effects that depend on the interior composition of the stars accumulate from roughly a few hundred Hz all the way up to merger, and not in a narrow band in the kHz range. Of course, these effects are much smaller at hundreds of Hz than at contact, however, the noise configuration that accumulates the most SNR when finite-size effects are non-negligible is the Zero-Detuned, High-Power one.

The above picture could, again, be altered if the merger was included. Numerical relativity studies~\cite{Bauswein:2009im,Stergioulas:2011gd,Hotokezaka:2011dh,Bauswein:2011tp,Bauswein:2012ya,Bernuzzi:2013rza,Hotokezaka:2013mm,Clark:2014wua,Bauswein:2014qla,Takami:2014zpa,Takami:2014tva,Hotokezaka:2015xka,Bauswein:2015yca,Bernuzzi:2015rla} have shown that the EoS leaves an important imprint in the merger and post-merger phases in the form of resonantlike features in the Fourier GW amplitude due to oscillations of hypermassive NSs. For such features that spike in a very narrow band, it may be that a high-frequency tuned noise curve would be best. As of today, it is unclear how much physical information could be extracted from such very high frequency features.

In our study we attempted to focus on physical questions of model selection as opposed to analyzing all possible EoSs constructed. There may exist other interesting physical features of NS interiors, other than the ones studied here, that might be worth investigating. To discover what other features would be interesting to measure, a stronger synergy between the GW and the nuclear physics community should be encouraged.   

\acknowledgments
We would like to thank Laura Sampson for many useful discussions. We would also 
like to thank Mark Alford, Jim Lattimer, Bennet Link, Madappa Prakash, Sanjay 
Reddy and Hajime Togashi for giving us valuable comments on the NS EoSs. We 
would like to thank Neha Gupta, Tetsuya Katayama, Benjamin Lackey, Yeunhwan Lim 
and Yasuo Yamamoto for kindly sharing with us their EoSs. K.C. acknowledges 
support from the Onassis Foundation. N.Y. acknowledges support from NSF CAREER Grant PHY-1250636. N.C. acknowledges support from the 
NSF Award PHY-1306702. A.K. is
supported by NSF CAREER Grant PHY-1055103.

\appendix

\section{EoS Classification}
\label{app-eos}

In this appendix, we classify various EoSs into several categories depending on their physics/compositions and methods used to derive them~\cite{Lattimer:2000nx,Read:2008iy}, and describe each class briefly. The basic properties of each class are summarized in Table~\ref{table:EoS-summary}.

\begin{table*}
\begin{centering}
\begin{tabular}{c|ccccc}
\hline
\hline
\noalign{\smallskip}
Class & EoS & Method/Model & Composition & Relativistic &  TNI  \\
\hline
\noalign{\smallskip}
Ia & AP4~\cite{APR} & variational & n, p, e, $\mu$& 1PN & Yes \\
Ia & GCR~\cite{Gandolfi:2011xu} & variational & n & (Yes) & Yes \\
Ib & SV~\cite{Lim:2013tqa} &  SHF & n, p, e, $\mu$& No & No \\
Ib & SGI~\cite{Lim:2013tqa}, SkI4~\cite{1995NuPhA.584..467R} &  SHF & n, p, e, $\mu$& No & Yes  \\
Ic & DBHF$^{(2)}$(A)~\cite{2014arXiv1410.7166K} & BHF & n, p, e, $\mu$& Yes & No \\
Ic & MPa~\cite{2014PhRvC..90d5805Y} & BHF & n, p, e, $\mu$& No & Yes  \\
Id &  G4~\cite{Lackey:2005tk}, GA-FSU2.1~\cite{2013JPhG...40b5203G} & RMF & n, p, e, $\mu$& Yes & No\\
\noalign{\smallskip}
\hline
\noalign{\smallskip}
IIb & SGI-YBZ6-S$\Lambda$$\Lambda$3~\cite{2014arXiv1412.5722L}, SkI4-YBZ6-S$\Lambda$$\Lambda$3~\cite{2014arXiv1412.5722L}, & SHF & n, p, e, $\mu$, H& No & Yes \\
IIc & NlY5KK$^{*}$~\cite{2014arXiv1410.7166K}, & BHF & n, p, e, $\mu$, H& No & No\\
IIc & MPaH~\cite{2014PhRvC..90d5805Y}, & BHF & n, p, e, $\mu$, H& No & Yes\\
IId & H4~\cite{Lackey:2005tk} & RMF & n, p, e, $\mu$, H& Yes & No\\
IIIb & SGI178~\cite{Lim:2013tqa} & SHF& n, p, e, $\mu$, K & No & Yes\\
IIIb & SV222~\cite{Lim:2013tqa} & SHF & n, p, e, $\mu$, K & No & No\\
IIId & GA-FSU2.1-180~\cite{2013JPhG...40b5203G} & RMF & n, p, e, $\mu$, K & Yes & No\\
IVa & ALF4~\cite{Alford:2004pf}, ALF5 & variational & n, p, e, $\mu$, $\pi$, Q & 1PN & Yes\\
IVa &  GCR-ALF & variational & n, Q & (Yes) & Yes\\
V & SQM3~\cite{SQM} & MIT bag & Q (u, d, s) & Yes & ---\\
\noalign{\smallskip}
\hline
\hline
\end{tabular}
\end{centering}
\caption{Properties of each EoS class. The upper half corresponds to the EoSs with normal matter only, while the lower half corresponds to those with exotic matter. The first column shows the classification number while the second column is an example EoSs within each class. The third column describes the method or model that is adopted to calculate each EoS. The fourth column shows the composition of matter, where H, K, $\pi$, Q refer to hyperon, kaon, pion and quark respectively. The fifth column presents whether relativistic effects have been taken into account (GCR effectively takes relativistic effects into account by assuming such effects have a similar density dependence to the short-range TNIs), while the final column shows whether three nucleon interactions are included. All the EoSs listed here are valid only at zero temperature. }
\label{table:EoS-summary}
\end{table*}

\subsection{Normal matter EoSs}

First, we focus on EoSs with \emph{normal} matter, namely neutrons, protons, electrons and muons.
The EoS can be calculated as follows. In the core region, the total energy density of the $\mathrm{npe\mu}$ matter is given by
\ba
\epsilon (n_n,n_p,n_e,n_\mu) &=& \epsilon_N (n_n,n_p) + n_n m_n c^2 \nn \\
& & + n_p m_p c^2  + \epsilon_e (n_e) \nn \\
 & & + \epsilon_\mu (n_\mu)\,, 
\ea
where $n_i$ and $m_i$ $(i=\mathrm{n,p,e,\mu})$ correspond to the number density 
and the mass of the $\mathrm{npe\mu}$ matter respectively. $\epsilon_N (n_n,n_p) 
$ represents the total nucleon contribution, calculated from different methods 
for treating the many-body problem with the effective nucleon nucleon interaction (NNI) 
model constructed based on existing experimental data. $\epsilon_e (n_e)$ and 
$\epsilon_\mu (n_\mu)$ are the energy densities of electrons and muons 
respectively, which one treats as free Fermi gases since Coulomb contributions 
are negligible compared to the kinetic energies. Equilibrium conditions on 
chemical potentials $\mu_i$ $(i=\mathrm{n,p,e,\mu})$ for npe$\mu$ matter with 
respect to weak interactions are given by $\mu_n = \mu_p + \mu_e$ and $\mu_\mu = 
\mu_e$ with $\mu_j = \partial \epsilon/\partial n_j$.
Together with charge neutrality, $n_p = n_e + n_\mu$, one can express $n_i$ with the baryon density $n = n_n + n_p$. This means that one has the total energy density $\epsilon$ only in terms of $n$. The pressure can be derived from the first law of thermodynamics given by
\be
\label{eq:1st-law}
p(n) = n^2 \frac{d}{dn} \left( \frac{\epsilon(n)}{n} \right)\,.
\ee

\subsubsection*{(Ia) Variational Chain Summation Method:  AP4, GCR}
\label{app:Ia}

To calculate the EoS in the nonrelativistic limit, we need, in principle, to derive the energy of each component of nuclear matter by solving the many-body Schr$\ddot{\mathrm{o}}$dinger equation. To minimize complexity, Pandharipande~\cite{Pandharipande:1971up} proposed to estimate the total energy $E$ of the system by numerically finding the correlation function in the wave function $\Psi$ that minimizes $E = (\Psi, H \Psi)/(\Psi, \Psi)$, where $(,)$ denotes the inner product and $H$ is the Hamiltonian.

This variational method with certain constraints on the correlation function reduces to solving a simplified Schr$\ddot{\mathrm{o}}$dinger equation with an interaction potential that depends on the baryon density $n$. Solving this equation and performing a cluster series expansion on the Hamiltonian to keep up to two-body clusters (namely, neglecting three-body clusters and higher), one obtains $E (n)$, which is the sum of the kinetic and potential energies. From this energy, one can easily derive the pressure using Eq.~\eqref{eq:1st-law}.

AP4~\cite{APR} uses the Argonne $v_{18}$ (AV18) NNI model~\cite{Wiringa:1994wb} and the Urbana IX (UIX) model of TNI~\cite{Pudliner:1995wk}. The former is obtained by fitting the proton-proton and proton-neutron scattering data from the Nijmegen group~\cite{Stoks:1993tb} and is expected to include all leading many-body correlation effects. 
AP4 further includes the relativistic 1PN boost correction to the AV18 model. 

GCR~\cite{Gandolfi:2011xu} uses the quantum Monte Carlo method~\cite{1999PhLB..446...99S} to systematically study the effect of TNIs on the EoS by varying the strength of such interactions and the range of the short-range TNI force. Due to the complexity of the method, the EoS is constructed for pure neutron matter. The authors use the same NNI model as AP4 and effectively take relativistic effects into account by assuming that such effects have a similar density dependence to TNIs, as shown in~\cite{APR}. Below $n < 0.08 \; \mathrm{fm}^{-3}$, the EoS is matched to the crust EoS~\cite{1971ApJ...170..299B,1973NuPhA.207..298N,SLy}.

\subsubsection*{(Ib) Skyrme-Hartree-Fock (SHF) models: \\ SV, SGI, SkI4}
\label{app:Ib}

SHF EoSs are constructed by using a Hartree-Fock approximation with effective, Skyrme type NNI models~\cite{1956PMag....1.1043S}.  The ground state energy of nucleon matter is given by $\langle \Psi_0 | \hat{H}_N  | \Psi_0\rangle$, where $\Psi_0$ is the ground state wave function that includes nucleon correlations and minimizes the energy, while $\hat{H}_N$ is the bare nuclear Hamiltonian including NNIs and TNIs. The mean-field scheme approximates this energy as $\langle \Phi_0 | \hat{H}_N^\mathrm{eff}  | \Phi_0\rangle = \int \mathcal{H}(\bm{r}) d^3 r$, where $\Phi_0$ is the Hartree-Fock wave function, $\hat{H}_N^\mathrm{eff}$ is the effective nuclear Hamiltonian and $\mathcal{H}$ is the Hamiltonian density. 
The standard Skyrme type model uses the parametrized Hamiltonian density~\cite{vanGiai:1981zz,1997NuPhA.627..710C,1998NuPhA.635..231C,Lim:2013tqa} including both local and nonlocal terms, density-dependent terms and a spin-orbit interaction. Such a model has up to 10 free parameters that are determined by fitting to experimental data. Apart from these ten parameters, $\mathcal{H}$ also depends on the neutron and proton local matter, kinetic and spin densities, which is given through the Hartree-Fock wave functions. 

The SV NNI model~\cite{1975NuPhA.238...29B}
uses five out of ten parameters to fit experimental data of the total binding energies and charge radii of magic nuclei (forming complete shells within nuclei), and it does not include three-body interactions. The SGI NNI model~\cite{vanGiai:1981zz}
improves previous models so that it predicts reasonable values for the incompressibility, spin and spin-isospin Landau parameters and pairing matrix elements.
The SkI4 NNI model~\cite{1995NuPhA.584..467R} includes an additional parameter in the spin-orbit term such that it can fit the measurement of charge isotope shifts in Sr and Pb isotopes.

\subsubsection*{(Ic) Brueckner-Hartree-Fock (BHF) models: \\ DBHF$^{(2)}(A)$, MPa}
\label{app:Ic}

The effective NNI in BHF models is obtained through the G-matrix $G(n;\omega)$ (where $\omega$ represents the unperturbed energy of the interacting nucleons) by solving the Bethe-Goldstone equation self-consistently~\cite{Baldo:1997ag}. Such an equation 
depends on the bare NNI potential and the single particle energy. The only parameters in the theory are those that appear in the former while in the nonrelativistic case, the latter is given by the sum of the kinetic energy and the single particle potential as $e(k) = k^2/2 m + U(k;n)$ with $m$ representing the bare nucleon mass. 
The Brueckner-Hartree-Fock (BHF) approximation for $U(k;n)$ is given by
\ba
U(k;n) = \sum_{k' \leq k_F} \langle kk' | G[n;e(k) + e(k')]| kk' \rangle_a\,, \nn \\
\ea
where $k_F$ represents the Fermi momentum. The subscript ``a'' refers to the antisymmetrization of the matrix element. In the BHF approximation, the energy per nucleon is given by~\cite{Baldo:1997ag}
\ba
\frac{E}{A} &=& \frac{3 k_F^2}{10 m} +  \frac{1}{2A} \sum_{k \leq k_F} U(k;n)\,,
\ea
with $A$ representing the nucleon number.

In the relativistic model, one introduces a strongly attractive scalar component and a repulsive vector component in the nucleon self-energy, which can be self-consistently obtained by solving a modified Thompson equation~\cite{1996ApJ...469..794E}. Such components add relativistic corrections to the nucleon mass and the single particle energy, which then modifies the G matrix and the energy per nucleon. The pressure is again obtained from the first law of thermodynamics, given in Eq.~\eqref{eq:1st-law}.

DBHF$^{(2)}$(A)~\cite{2014arXiv1410.7166K} is constructed by solving the coupled Bethe-Saltpeter equations in the rest frame of nuclear matter using the Bonn A potential~\cite{Brockmann:1990cn} within the ``reference spectrum'' approximation, which assumes that the values of self-energies are frozen at the Fermi momentum at each total baryon density. Nucleons interact through exchanges of two scalar ($\sigma$, $\delta$), two vector ($\omega$, $\rho$) and two pseudoscalar ($\eta$, $\pi$) mesons. 

MPa~\cite{2014PhRvC..90d5805Y}\footnote{In this paper, we refer to the MPa EoS without hyperons as ``MPa'' and that with hyperons as ``MPaH''.} is constructed within the nonrelativistic BHF model with three-nucleon interactions taken into account. The NNI is described by the extended soft core model~\cite{Rijken:2010zzb} which explicitly includes two-meson and meson-pair exchanges. TNIs have two parts, repulsive and attractive. In particular, the former is important to make the EoS stiffer so that it can support NSs with a mass larger than 2$M_\odot$. The three-nucleon repulsive (TNR) part is modeled by the multi-Pomeron exchange potential including the triple and quartic Pomeron exchange, where the Pomeron is related to an even number of gluon exchanges. The strength of the TNR part is determined from {}$^{16}$O+{}$^{16}$O elastic scattering experiments~\cite{Furumoto:2009zza,Furumoto:2009zz}. The three-nucleon attractive part is added such that it reproduces the nuclear saturation property~\cite{1977PhRvD..16..466B}. Having such an interaction model at hand, one can calculate the G-matrix, and in turn the EoS with chemical equilibrium conditions, charge neutrality and baryon number conservation.

\subsubsection*{(Id) Relativistic mean field (RMF) theory: \\ G4, GA-FSU2.1}
\label{app:Id}

Reference~\cite{Lackey:2005tk} constructs the G4 EoS, where the authors model the low-energy strong nuclear interaction as the leading exchange of mesons and baryons. One starts with the construction of a relativistically covariant Lagrangian, which includes free baryons, leptons, $\sigma$, $\omega$ and $\rho$ mesons, the tree level meson-baryon interactions and perturbative self-interactions for the $\sigma$ meson. The theory is a phenomenological low-energy effective field theory. Such a theory has a small number of parameters, which are determined from experiments. The fields are replaced by their mean values, assuming that the bulk matter is static and homogeneous. The derivatives of the fields with respect to time and space also vanish, which simplifies the Euler-Lagrange equations. Such equations are combined with those of charge and baryon number conservations and  $\beta$-equilibrium, to yield the Fermi momenta and meson fields as functions of the baryon density. One then uses these solutions to derive the energy and pressure.

Gupta and Arumugam~\cite{2013JPhG...40b5203G} constructed an EoS using the 
effective field theory-motivated RMF (E-RMF) model~\cite{Furnstahl:1996wv}, 
which has a few additional couplings on top of the standard RMF models. The 
effective Lagrangian in the E-RMF model is obtained in a power series of fields 
and their derivatives to a certain order, and all the nonrenormalizable 
couplings are made to be consistent with the symmetries of quantum chromodynamics (QCD). The E-RMF model can 
simultaneously explain finite nuclei and infinite matter~\cite{Arumugam:2004ys}. 
In this paper, we use the EoS in~\cite{2013JPhG...40b5203G} with the FSU2.1 
parameters for the nucleon-meson coupling constants, which we call the GA-FSU2.1 
EoS.

\subsection{EoSs with exotic matter}

Let us now consider EoSs that also include exotic matter, like hyperons, kaons and quarks. 

\subsubsection*{(IIb) Hyperons with SHF Models: \\ SGI-YBZ6-S$\Lambda$$\Lambda$3, SkI4-YBZ6-S}
\label{app:IIb}

Reference~\cite{2014arXiv1412.5722L} included the $\Lambda\Lambda$ hyperon interactions into a few Skyrme-type EoSs, in particular, SGI and SkI4. 
As in constructing SHF models with normal matter components, Ref.~\cite{2014arXiv1412.5722L} parametrized the Hamiltonian density for a $\Lambda$ hyperon~\cite{Lanskoy:1997xq} including the $\Lambda \Lambda$ interaction~\cite{Lanskoi:1998gn}, whose parameters are again determined from experiments. Several sets for these parameters exist~\cite{Lanskoy:1997xq,Yamamoto:1988qz,Fernandez:1989zw,Guleria:2011kk}, including the YBZ6 model~\cite{Yamamoto:1988qz} for the parameters associated with the N$\Lambda$ interaction and the S$\Lambda\Lambda$3 model~\cite{Lanskoi:1998gn} for the $\Lambda\Lambda$ interaction.
On top of these Skyrme-type interaction potentials, the authors in~\cite{2014arXiv1412.5722L} introduced a finite-range force model for the $\Lambda\Lambda$ interaction. This is because the binding energies calculated from the Skyrme-force models are not fully consistent with experimental data. Parameters in such a non-Skyrme-type interaction are determined from the measured binding energy of ${}_{\Lambda\Lambda}^6$He~\cite{Hiyama:2002yj}.

Based on these interaction potentials, the EoSs are calculated under $\beta$-equilibrium conditions, charge neutrality and baryon number conservation. 
The EoS constructed with the SGI (SkI4) NNI model, the YBZ6 model for the N$\Lambda$ interaction and the S$\Lambda\Lambda$3 model for the $\Lambda\Lambda$ interaction is called the SGI-YBZ6-S$\Lambda\Lambda$3 (SkI4-YBZ6-S$\Lambda\Lambda$3) EoS.

\subsubsection*{(IIc) Hyperons with BHF models: \\ NlY5KK$^*$, MPaH}
\label{app:IIc}

The NlY5KK$^*$ EoS~\cite{2014arXiv1410.7166K} is constructed by extending DBHF$^{(2)}$(A) to include the effect of hyperons ($\Lambda$, $\Sigma^-$ and $\Xi^-$) with the vector K$^*$ and pseudoscalar $K$ mesons which induce the baryon-exchange and baryon-transition processes. The hyperon-meson coupling constants are determined from SU(6) symmetry~\cite{1994NuPhA.570..543R}. 

The MPaH EoS is an extension of the MPa EoS that includes hyperons ($\Lambda$ and $\Sigma^-$). This EoS is constructed under the assumption that TNR-like repulsive interactions among three nucleons work universally for HNN, HHN, HHH (as well as NNN), where N and H refer to a nucleon and hyperon respectively. 

\subsubsection*{(IId) Hyperons with RMF theory: H4}
\label{app:IId}

Reference~\cite{Lackey:2005tk} constructs EoSs with hyperons using RMF theory by adding such effects on top of the G4 EoS.
Hyperons are produced at roughly twice nuclear density. The most relevant hyperons are $\Lambda$ and $\Sigma^-$ hyperons, which have the smallest masses. The hyperon-meson couplings are taken to be the same among all hyperons and are smaller than those between nucleons and mesons. The H4 EoS is constructed by setting $(\tilde K,m^*/m,x_\sigma)=(300\mathrm{MeV}, 0.70, 0.72)$, where $\tilde K$, $m$, $m^*$ and $x_\sigma$ represent the incompressibility, the bare nucleon mass, the effective nucleon mass and the ratio between the hyperon-meson and nucleon-meson couplings for the $\sigma$ meson respectively. 

\subsubsection*{(IIIb) Kaon condensates with SHF models: \\ SV222, SGI178}
\label{app:IIIb}

For a given density, all the parameters in the nucleon and lepton Hamiltonian densities are determined from the $\beta$-equilibrium and charge neutrality conditions, and the properties of finite nuclei~\cite{Kortelainen:2010hv}. The kaon Hamiltonian density consists of the kinetic and mass terms of the kaon and the kaon-nucleon interactions, adopting a SU(3) nonlinear chiral model first proposed in~\cite{Kaplan:1986yq}. One of the parameters in the kaon Hamiltonian density, which corresponds to the magnitude of the expectation value of the kaon condensate, is determined by minimizing the total energy density. The pressure is derived from Eq.~\eqref{eq:1st-law}. The three remaining parameters, $a_1$, $a_2$ and $a_3$ in the kaon Hamiltonian density are chosen as $a_1 m_s = -67$MeV, $a_2 m_s = 134$MeV~\cite{Prakash:1996xs} and $a_3 m_s = -222$MeV for an SV EoS (SV222) and $a_3 m_s = -178$MeV for an SGI EoS (SGI178), with $m_s$ representing the mass of the strange quark.

\subsubsection*{(IIId) Kaon condensates with RMF theory: \\ GA-FSU2.1-180}
\label{app:IIIb}

Reference~\cite{2013JPhG...40b5203G} also constructed an EoS based on the E-RMF theory with kaon (and antikaon) condensates ($K^-$ and $\bar K^0$). On top of the effective Lagrangian for nuclear matter, the following Lagrangian for kaon condensates are added: $ \mathcal{L}_K = D_\mu^* K^* D^\mu K - \bar m_K^2 K^* K$, where $K = K^-$ or $\bar K^0$, $\bar m_K$ represents the effective mass of kaons that depends on the scalar meson fields and $D_\mu$ is the differential operator that depends on the vector and isovector meson fields. The presence of kaon condensates modifies the expectation values of the meson fields that are used to replace the fields themselves in the Lagrangian. The total energy density is given by $\epsilon = \epsilon_N + \epsilon_{K}$ with $\epsilon_K = \bar m_K (n_{K^-} + n_{\bar K^0})$,
where $\epsilon_N$ and $\epsilon_K$ represent the energy density of the nucleon phase~\cite{Gupta:2012zza} and kaon condensates respectively, while $n_{K^-}$ and $n_{\bar K^0}$ are the number density of $K^-$ and $\bar K^0$. On the other hand, the expression for pressure is unaffected by the presence of kaon condensates~\cite{Glendenning:1997ak}. In this paper, we use the EoS with the FSU2.1 parameters and the optical potential of a single kaon in infinite matter as $U_K = -180$MeV, which we call the GA-FSU2.1-180 EoS.

\subsubsection*{(IVa) Hybrid nuclear and quark matter \\ with variational method: ALF4, ALF5, GCR-ALF}
\label{app:IVa}

Reference~\cite{Alford:2004pf} constructs the hybrid EoS of the ALF family for nuclear and quark matter. The former is modeled using the APR EoS described above. The EoS in the low density region is modeled by the standard tabulated EoS of~\cite{1971ApJ...170..299B,1973NuPhA.207..298N}. The quark matter EoS is based on a physical model, which takes into account both phases of normal unpaired quark matter and color-flavor-locked (CFL) quark matter. In the latter, quarks form Cooper pairs and one-to-one correspondence arises between three color pairs and three flavor pairs. The phase with the lower free energy is favored. 

Neglecting perturbative QCD corrections, the free energy (or the grand 
potential) density of unpaired quark matter $\Omega_\mathrm{unp} (\mu_q)$ (with 
$\mu_q$ representing the chemical potential for a quark) contains the kinetic 
contribution from a degenerate free gas of three colors of relativistic quarks 
and the negative vacuum pressure represented by the bag constant $B$. Such a 
constant depends on the transition density $n_c$ at which the quark matter EoS 
is stitched to the nuclear matter one. The authors also introduced a QCD 
inspired term~\cite{Fraga:2001id}, which is proportional to $(1-c) \mu_q^4$. 
Here, $c$ is the QCD correction parameter where $c=0$ corresponds to 
noninteracting, free quarks. Reference~\cite{Fraga:2001id} carried out a 
perturbative calculation and found $c \sim 0.37$, although higher order 
corrections in the strong coupling constant may be important for hybrid stars. 
$c$ determines the maximum mass of a star with a hybrid EoS and such a mass 
becomes larger as one increases $c$.

Regarding CFL quark matter, the free energy can be decomposed into three parts:
\ba
\label{eq:CFL}
\Omega_\mathrm{CFL} (\mu_q,\mu_e) &=& \Omega_\mathrm{CFL}^\mathrm{neutral} (\mu_q) + \Omega_\mathrm{CFL}^\mathrm{\pi} (\mu_q,\mu_e) \nn \\
& & + \Omega^\mathrm{leptons} (\mu_e)\,.
\ea
The first term, $\Omega_\mathrm{CFL}^\mathrm{neutral} (\mu_q)$, denotes the contribution from the neutral CFL phase. The second term represents the contribution from pion condensates. Such condensates arise when the electron chemical potential exceeds the mass of the $\pi^{-}$ meson, the lightest negatively charged meson in the CFL phase~\cite{2002NuPhA.697..802B,Kaplan:2001qk,Kryjevski:2004cw}. Finally, the third term is the contribution from electrons and muons.
The quark matter EoS is then constructed by using the thermodynamic relations:
\be
\label{eq:EoS-quark}
p = - \Omega\,, \quad \epsilon = \Omega - \frac{d\Omega}{d \ln \mu_q}\,.
\ee
The matching between the quark and nuclear matter EoS is carried out by imposing global charge neutrality and pressure balance conditions between the two matter phases~\cite{Glendenning:1992vb,Alford:2001zr}, where the latter condition specifies $\mu_e$ in terms of $\mu_q$.

Selected ALF EoSs (ALF1--4) were used in~\cite{Read:2008iy}. In particular, ALF4 
has $(n_c,c)=(4.5n_0,0.3)$ where $n_c$ and $n_0$ represent the transition and 
saturation densities respectively. The m-R relation for the ALF4 EoS is based on 
the EoS data constructed by the authors in~\cite{Alford:2004pf} and is slightly 
different from that used in~\cite{Read:2008iy}. It seems that the latter was 
constructed by artificially stitching the AP4 EoS to the data constructed 
in~\cite{Alford:2004pf}, but such a procedure changes some part of the mixed phase 
into the pure nuclear matter phase, which is energetically more disfavored than 
the mixed phase (so this artificial pure nuclear matter phase is unstable). 

Reference~\cite{Read:2008iy} claims that ALF2 can support a 2$M_\odot$ NS (and has a typical NS radius of $\sim 13$km) but we do not use this EoS due to the following uncertainty. According to~\cite{Read:2008iy}, this EoS has parameters $(n_c,c)=(3n_0,0.3)$ but Fig.~6 of~\cite{Alford:2004pf} shows that the maximum mass with such a choice of parametrization is below 2$M_\odot$ and a typical radius is $\sim 11$km. The ALF2 EoS data used in~\cite{Read:2008iy} has a transition of ``AP4 $\to$ CFL $\to$ mixed phase $\to$ CFL'' as one increases the density, although it is not clear if such a transition is physically reasonable (ALF1, 3 and 4 all have a transition of ``AP4 $\to$ mixed phase $\to$ CFL''). Instead, following~\cite{Alford:2004pf}, we constructed a new ALF EoS (ALF5)  with $(n_c,c)=(2n_0,0.4)$ that can support a 2$M_\odot$ NS.   

We also constructed new hybrid EoSs (GCR-ALF) by stitching the nuclear matter GCR EoSs~\cite{Gandolfi:2011xu} with the quark matter EoSs in~\cite{Alford:2004pf}. Instead of imposing the global charge neutrality condition, we impose charge neutrality on each nuclear matter and quark matter (CFL) phase~\cite{Alford:2004pf}. The charge neutrality on the CFL phase leads to the absence of electrons (and muons and pion condensates), thus $\mu_e=0$ and $ \Omega_\mathrm{CFL}^\mathrm{\pi} =0 = \Omega^\mathrm{leptons} $ in Eq.~\eqref{eq:CFL}.  The nuclear and quark matter EoSs are matched at a chemical potential where the pressure between the two phases becomes identical and such a construction leads to the absence of the mixed phase~\cite{Alford:2004pf}. As an example, Fig.~\ref{fig:ALF2AP4} presents the m-R relation for a GCR-ALF EoS with a symmetry energy of $E_\mathrm{sym}=33.8$MeV in the nuclear matter EoS (with the crust EoS in~\cite{SLy}) and $(n_c,c)=(2n_0,0.35)$ in the quark matter EoS, together with the one for the corresponding GCR EoS. 
We have also constructed GCR-ALF EoSs with other parametrizations, but found that the difference in the m-R relation between GCR and GCR-ALF EoSs is typically smaller than that shown in Fig~.\ref{fig:ALF2AP4}.

\subsubsection*{(V) Strange quark matter: SQM3}
\label{app:V}

SQM EoSs are constructed from the MIT bag model. When one can neglect the mass of quarks, the free energy density is given by $\Omega = C_q \mu^4 + B$ with $C_q$ representing a constant. From Eq.~\eqref{eq:EoS-quark}, the EoS for quark matter becomes $p = 3 (\epsilon - 4B)$. The prefactor ``3'' changes if one includes the effect of the strange quark mass $m_s$.
SQM3 assumes $m_s=50$MeV and $B=57.39$
MeV/fm$^3$, where the latter corresponds to the minimum bag constant allowed from the stability of neutrons with respect to a spontaneous fusion into strangelets~\cite{Stergioulas:2003yp}.

\section{Bayes Factors and Prior Cutoffs}
\label{cutoffs}

The NSNS Optimized noise curve has higher noise levels at high frequencies than the Zero-Detuned, High-Power one. Therefore, as far as EoS determination is concerned, the likelihood calculated with a NSNS Opt.~noise curve does not contain a lot of information in the high-frequency part of the waveform where finite-size effects become important. Calculating BFs in a regime where it is not the likelihood, but the prior that dominates the results can lead to an interesting and rather counterintuitive effect: BFs that initially decrease with increasing SNR. In other words, as the signal strength increases, the correct model is preferred less and less. 
\begin{figure}[t]
\begin{center}
\includegraphics[width=\columnwidth,clip=true]{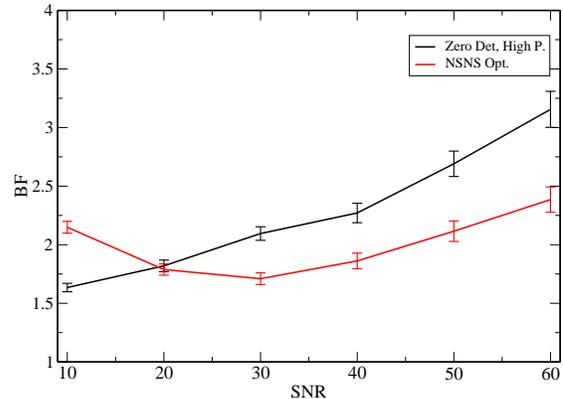}
\caption{\label{fig:BFnoise}  BF in favor of SV compared to SV222 for the system with masses $(1.95,1.9)$ with the Zero-Detuned, High-Power sensitivity curve (black) and the NSNS Optimized curve (red). For high SNR values, where the likelihood dominates, the detuning curve gives higher BFs. For lower SNR values with the NSNS Opt. curve, we encounter the counterintuitive effect of decreasing BFs with increasing SNR values. }
\end{center}
\end{figure}
\begin{figure}[t]
\begin{center}
\includegraphics[width=\columnwidth,clip=true]{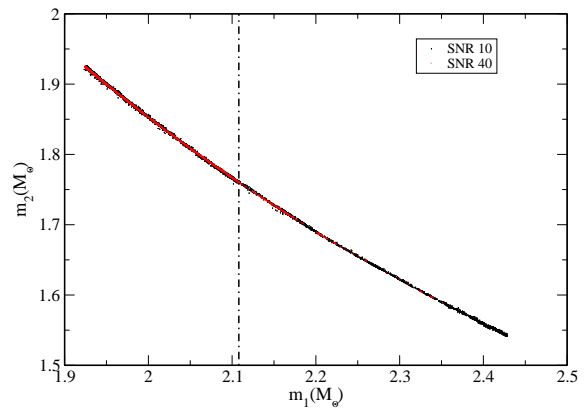}
\caption{\label{fig:m1m2SNR}  $2$D Scatter plot in the $m_1-m_2$ plane for SNR 10 (black) and 40 (red) with the NSNS Opt. sensitivity curve and with the same system as in Fig.~\ref{fig:BFnoise}. The vertical line denotes the maximum SV222 mass. Any points on the right of this line necessarily correspond to the SV model.}
\end{center}
\end{figure}

Figure~\ref{fig:BFnoise} shows the BF in favor of SV compared to SV222 for the $(1.95,1.9)M_{\odot}$ system calculated with the Zero-Det., High-P.~curve (black) and the NSNS Opt. curve (red). The black line is the same as the black dashed line of Fig.~\ref{fig:bf_kaons}. The red line presents some rather interesting behavior. At low SNR, the BF decreases with the SNR, while after SNR $=40$ it starts increasing, like one would expect. In order to understand this effect, consider Fig.~\ref{fig:m1m2SNR}, where we plot the $2-$D scatter plot of the chain points in the $m_1-m_2$ plane for the system used in Fig.~\ref{fig:BFnoise}, analyzed with NSNS Opt., and SNR 10 (black dots) and 40 (red dots). The vertical line indicates the maximum mass the competing model SV222 can support. This maximum mass is a hard cutoff on the prior mass of SV222; no system with masses higher than the maximum mass can be produced by SV222, and the model reduces to noise. This effectively means that SV222 has no posterior weight in that region, and all these points correspond to SV.

As the SNR increases the posterior width decreases. At SNR $=40$ there are very few points above the mass cutoff. That means that with increasing SNR SV222 suffers less and less from this cutoff in its mass prior. This explains why the BF in favor of SV decreases as we increase the SNR. Of course, at some values of the SNR the differences between the models will start dominating over the prior cutoff, and the BF will again start increasing with the SNR. Figure~\ref{fig:BFnoise} shows that in this case this is true for SNR $> 40$.

\section{Toy Model}
\label{toymodel}

\begin{figure}[t]
\begin{center}
\includegraphics[width=\columnwidth,clip=true]{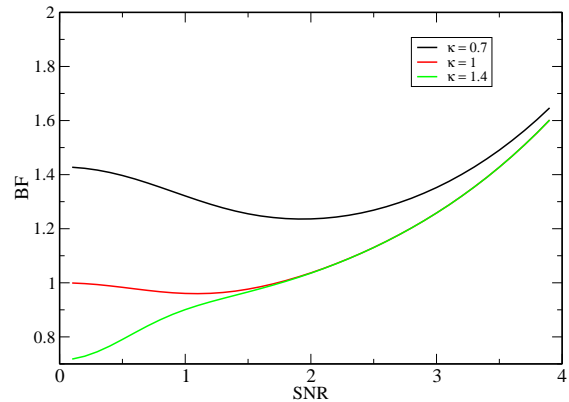}
\caption{\label{fig:BFtoy}  BF in favor of $h_1$ as a function of the SNR for $\kappa=1$ (red), $\kappa=0.7$ (black), and $\kappa=1.4$ (green). The effect of the BF decreasing with increasing SNR is present in both $\kappa=1$ and $\kappa=0.7$ cases, but it is more pronounced in the case where the two models have a different parameter range. As the SNR increases all three lines tend to coincide. This is because the likelihood becomes more and more peaked, and at some value of the SNR the limits of integration stop affecting the integral of the likelihood.}
\end{center}
\end{figure}
\begin{figure}[t]
\begin{center}
\includegraphics[width=\columnwidth,clip=true]{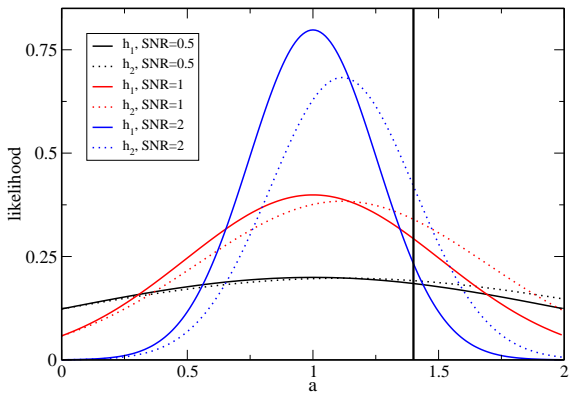}
\caption{\label{fig:likelihood}  Likelihood for the correct model (solid lines) and the wrong model (dotted lines) for different values of the SNR. The black vertical line shows the cutoff in the parameter prior range when $\kappa=0.7$. The likelihood for the wrong model does not peak at the injected value of $a=1$ and it is not a Gaussian.}
\end{center}
\end{figure}

To demonstrate the rather counterintuitive effect of BFs that favor the wrong model, or decrease with the SNR, we construct a simple toy problem, thus obtaining a more robust explanation of these effects than the 2-D scatter plots presented in the previous appendix. Imagine we receive N data points from a very simple signal that obeys $d(f)=f$ and we try to match it with two competing $1-$D models $h_1(f)= a f$ and $h_2(f)=a f^{1.5}$, where $a$ is the parameter of the models. The likelihood for model $i$ is
\be
L_i = \frac{1}{\sqrt{2 \pi} \sigma} \exp{\left\{-\sum^N  \frac{[d(f)-h_i(f)]^2}{2 \sigma^2}\right\}}.
\ee
The parameter $\sigma$ is the standard deviation of the data; in GW language, it is 1/SNR. Clearly $h_1$ is the correct model and $L_1$ is maximized when $a=1$. On the other hand, $h_2$ cannot fit the signal perfectly; $L_2$ is maximized for $a=1.12$ with a residual that depends on $\sigma$. 

Now imagine that the two models have different prior ranges for the parameter $a$. In the case of $h_1$ we have $a\in(0,2)$, while $h_2$ allows $a\in(0,2\kappa)$, where $\kappa$ is an arbitrary number. The evidence for each model is proportional to its likelihood integrated over the range of parameter $a$, while the BF in favor of $h_1$ (the correct model) is the ratio of the two evidences. Figure~\ref{fig:BFtoy} shows the BF in favor of $h_1$ as a function of the SNR $(=1/\sigma)$ for three different values of $\kappa$. The case $\kappa=1$ corresponds to $2$ models with the same parameter prior range, while $\kappa=0.7$ and $\kappa=1.4$ are similar to comparing EoS models with different maximum allowed mass. More specifically, $\kappa = 0.7$ corresponds to the case where the wrong model has the smaller prior range (the right panels of Figs.~\ref{fig:bf_kaons},~\ref{fig:bf_hyperons}, and Fig.~\ref{fig:BFnoise}), while $\kappa=1.4$ corresponds to the case where it is the correct model that has the smaller prior range (the left panels of Figs.~\ref{fig:bf_kaons} and~\ref{fig:bf_hyperons}).

The $\kappa=1.4$ case is easier to understand. The likelihood of the wrong model $h_2$ is integrated over a larger parameter region $a\in(0,2.8)$ than the correct model where $a\in(0,2)$. When the SNR is low and the models are not very different from each other, this can lead to a larger evidence for the wrong model. As the signal strength increases the differences between the two models will start dominating the evidence and the correct model will end up being preferred. Indeed we find that for SNR $\gtrsim2$ the BF favors the correct model $h_1$.

Both $\kappa = 0.7$ and $\kappa=1$ lead to BFs that decrease with increasing SNR, however, this effect is more pronounced in the $\kappa=0.7$ case and it extends to higher values of SNR making it easier to identify. When $\kappa=1$, one might expect the Laplace approximation to the evidence to be reliable since the posterior width on $a$ is smaller than the prior range. However, this is not the case: the likelihood for $h_2$ is not a Gaussian, and it is this small deviation from Gaussianity that we see as a decreasing BF. To visualize this effect, in Fig.~\ref{fig:likelihood} we plot the likelihood for $h_1$ and $h_2$ for different values of SNR. For low SNR values the likelihoods are essentially the same and BF $=1$. However, as the SNR increases the area under the red dotted line is larger that the area under the red solid line, leading to a small decrease in the BF in favor of $h_1$. Clearly at sufficiently high SNR the correct model will prevail and the BF will start increasing in favor of $h_1$.

When $\kappa=0.7$ the effect of BFs that decrease with SNR is stronger and persists for higher values of SNR. Revisit Fig.~\ref{fig:likelihood} and keep in mind now that the evidence for the wrong model is obtained by integrating the likelihood up to the black vertical line. As the SNR increases, the likelihood for $h_2$ is more peaked on the left of the vertical line, which means that the area permissible by the prior cutoff increases. This leads to an increase of the evidence of $h_2$ and a BF in favor of $h_1$ that decreases. Clearly at some point the differences between the models will overcome this effect, and the correct model will prevail. For our toy model this happens when SNR $\gtrsim2$.

\bibliography{review}
\end{document}